\documentclass[%
 amsmath,amssymb,amsfonts,
 aps,
rmp, a4paper
,twocolumn,hidelinks
]{revtex4-2}
\usepackage{epsfig}
\usepackage{epsfig}
\usepackage{graphicx}
\usepackage{dcolumn}
\usepackage{bm}
\usepackage{hyperref}
\usepackage{xcolor}
\usepackage{mathrsfs}
\usepackage{siunitx}
\usepackage{comment}
\usepackage{cancel}
\usepackage{tensor}
\usepackage[toc,page]{appendix}
\newcommand{\nocontentsline}[3]{}
\newcommand{\tocless}[2]{\bgroup\let\addcontentsline=\nocontentsline#1{#2}\egroup}
\usepackage{physics}
\usepackage{comment}
\usepackage{siunitx}
\newcommand{\cost}{\mathcal{C}}
\newcommand{\basis}{\mathcal{B}}

\newcommand{\treversal}{\hat{T}}

\usepackage[normalem]{ulem}
\newcommand{\bR}{\mathbf{R}}

\newcommand{\dbR}{\Delta\mathbf{R}}
\newcommand{\bravais}{\mathfrak{B}}

\newcommand{\bravaisbulk}{\mathfrak{B}_{\scriptscriptstyle \text{bulk}}}
\newcommand{\bulk}{\mathcal{B}}

\newcommand{\fourier}{\mathfrak{F}}
\newcommand{\Eig}{\mathfrak{E}}

\newcommand{\br}{\mathbf{r}}
\newcommand{\tE}{\tilde{\mathcal{E}}}

\newcommand{\N}{\mathcal{N}}

\newcommand{\hE}{\hat{\mathcal{E}}}
\newcommand{\E}{\mathcal{H}}
\newcommand{\DE}{\Delta\mathcal{H}}
\newcommand{\barq}{\bar{q}}
\newcommand{\K}{\mathcal{K}}

\newcommand{\tilQ}{\tilde{Q}}

\newcommand{\Q}{\mathcal{Q}}
\newcommand{\tiQ}{\tilde{\mathcal{Q}}}

\newcommand{\bw}{\mathbf{w}}

\newcommand{\be}{\mathbf{e}}

\newcommand{\complex}{\mathbb{C}}
\newcommand{\templim}{${T\to 0\;}$}

\newcommand{\bu}{\mathbf{u}}
\newcommand{\bcdot}{\boldsymbol{\cdot}}
\newcommand{\dbu}{\dot{\mathbf{u}}}

\newcommand{\bP}{\mathbf{P}}

\newcommand{\ba}{\mathbf{a}}

\newcommand{\bv}{\mathbf{v}}
\newcommand{\bx}{\mathbf{x}}

\newcommand{\bpsi}{\boldsymbol{\psi}}

\usepackage{mathtools}

\newcommand{\dbpsi}{\boldsymbol{\dot{\psi}}}

\newcommand{\kket}[1]{|#1)}

\newcommand{\dpsi}{\dot{\psi}}

\newcommand{\bbra}[1]{(#1|}
\newcommand{\bbraket}[2]{(#1|#2)}

\newcommand{\ddyad}[2]{\kket{#1}\bbra{#2}}

\newcommand{\tHess}{\boldsymbol{\tilde{\mathsf{H}}}}
\newcommand{\Dyn}{\boldsymbol{\mathsf{D}}}
\newcommand{\tDyn}{\boldsymbol{\tilde{\mathsf{D}}}}

\newcommand{\bk}{\mathbf{k}}

\newcommand{\bq}{\mathbf{k}}

\newcommand{\bbq}{\bar{\bq}}
\newcommand{\bbk}{\bar{\bk}}

\newcommand{\eig}{\epsilon}
\newcommand{\neig}{\varepsilon}

\newcommand{\tpsi}{\tilde{\psi}}
\newcommand{\hpsi}{\hat{\psi}}
\newcommand{\btpsi}{\boldsymbol{\tilde{\psi}}}

\newcommand{\Mss}{\boldsymbol{\Upsilon}}

\newcommand{\pol}{\pi}

\newcommand{\bomega}{\bar{\omega}}

\newcommand{\bG}{\mathbf{G}}
\DeclareMathAlphabet\mathbfcal{OMS}{cmsy}{b}{n}

\newcommand{\bQ}{\mathbf{Q}}

\newcommand{\volume}{\abs{\Omega}}

\newcommand{\bb}{\mathbf{b}}
\newcommand{\realone}{\mathbb{R}}
\newcommand{\realpos}{\mathbb{R}^+}

\newcommand{\integer}{\mathbb{Z}}

\newcommand{\integernonneg}{\mathbb{Z}^+_0}
\newcommand{\bz}{\rotatebox[origin=c]{180}{$\Omega$}}
\newcommand{\bza}{\hat{\bz}}
\newcommand{\bzsub}{\rotatebox[origin=c]{180}{{${\scriptstyle \Omega}$}}}
\newcommand{\bzasub}{\hat{\bzsub}} 
\newcommand{\bzamin}{\hat{\bz}_{\text{min}}}
\newcommand{\bzam}{\hat{\bz}_-}
\newcommand{\bzapm}{\hat{\bz}_\pm}
\newcommand{\nlin}{m}

\newcommand{\hC}{\hat{C}}
\newcommand{\bzero}{\mathbf{0}}

\makeatletter
\newcommand{\subalign}[1]{%
  \vbox{%
    \Let@ \restore@math@cr \default@tag
    \baselineskip\fontdimen10 \scriptfont\tw@
    \advance\baselineskip\fontdimen12 \scriptfont\tw@
    \lineskip\thr@@\fontdimen8 \scriptfont\thr@@
    \lineskiplimit\lineskip
    \ialign{\hfil$\m@th\scriptstyle##$&$\m@th\scriptstyle{}##$&$\m@th\scriptstyle{}##$\hfil\crcr
      #1\crcr
    }%
  }%
}
\makeatother
\usepackage[T1]{fontenc}

\newcommand{\wk}{${(\bk,\omega)}$}

\newcommand{\tmax}{\mathcal{T}}

\newcommand{\maxomega}{\mathcal{W}}
\newcommand{\wannier}{w}

\begin{document}

\preprint{APS/123-QED}

\title{General theory of lattice dynamics}
\title{Theory of lattice dynamics}
\title{The wave theory of lattice dynamics}
\title{Wave theory of lattice dynamics and its continuum limit}
\title{Energy is correlation}
\title{Wave theory of lattice dynamics}

\author{Paul Tangney}%
\email{p.tangney@imperial.ac.uk}
\affiliation{%
Department of Materials and Department of Physics, 
Imperial College London
}%


\date{\today}

\begin{abstract}
I present the mathematical structure of classical phonon theory in a general form, which
emphasizes the wave natures of phonons, and which can serve as a robust foundation for further
development of the theory of strongly interacting phonons. 
I also show that the Fourier transform (FT) of the mass-weighted velocity-velocity correlation function (mVVCF)
is {\em exactly} the distribution of the classical kinetic energy among frequencies and wavevectors.
Because this result is classically exact, it is general: It is as valid, theoretically, for a liquid or a molecule in a non-thermal non-stationary state as it is
for a crystal at thermal equilibrium at a low temperature.
Therefore, as well as being of fundamental importance to physical theory, this result implies
that calculating the FT of the mVVCF from atomistic simulations is a much more powerful computational tool than it is believed to be.
Existing theory shows only that the FT of the mVVCF is proportional to the vibrational density of states at thermal equilibrium, and under
the simplifying assumption that the number of available vibrational states is equal to the number of degrees of freedom.
\end{abstract}

\maketitle
\onecolumngrid
\tableofcontents{}
\vspace{2cm}
\twocolumngrid

\section{Introduction}
\label{section:theory}
One purpose of this article is to provide a careful account of how
the microstate of a hot crystal can be expressed in mathematical
forms that simplify the study of vibrations
in crystals by atomistic simulations.
It is surprisingly difficult to find a detailed and complete account 
of the mathematical structure of phonon theory
in existing literature: Most introductory textbooks make too many simplifying
assumptions for the requirements of those simulating spectra atomistically, 
whereas more advanced ones tend to build the theory with 
the quantum-mechanical many body theory of phonons in mind~\cite{Born_and_Huang,wallace,Jones_and_March1,Ashcroft_and_Mermin,kittel,ibach_and_luth,cohen_louie_2016}.

Quantum mechanical derivations of phonon theory 
usually treat phonons as quasiparticles that scatter from one another.
However phonons can also be treated as lattice waves
whose interactions are continuous, with energy and
momentum being exchanged everywhere in space at all times.
These two approaches, or emphases, rest on the same physical
assumptions, and are therefore equivalent; but
for many purposes, such as interpreting the results of atomistic simulations, 
the lattice wave picture is more useful.
Therefore, in order to relate the various methods of atomistically simulating
vibrational spectra to one another, we need a classical description 
of the crystal's structure and dynamics, from which all of these 
methods may be derived.

A second purpose of this article is to derive an expression 
for the distribution of a classical crystal's kinetic energy among
wavevectors (${\bk}$) and angular frequencies ($\omega$). I refer
to the set of all points \wk, which specify a wavevector and a frequency, as
{\em reciprocal spacetime}.
It is well known that, at thermal equilibrium and in the low temperature ($T$) limit, 
the distribution of kinetic energy in reciprocal spacetime is the Fourier transform of the atoms' mass-weighted velocity-velocity correlation function (mVVCF).
However, all derivations of this result that I am aware of make use of the equipartition theorem 
and assume that the number of available vibrational states is equal to the total number of degrees
of freedom. For example, in a crystal it is usually assumed that 
energy is localized at the points \wk~that identify the frequencies and wavevectors
of the crystal's normal modes of vibration.

I prove that the Fourier transform of the mVVCF is exactly the distribution
of kinetic energy in reciprocal spacetime.
I prove this result using basis sets that simplify the study of crystals, because that is my focus. However, because these basis sets
are complete, the result is general. Therefore it is likely to useful in domains other than materials science.
In the context of atomistic simulations, this result means that the reciprocal spacetime kinetic energy distribution can be calculated
from the mVVCF at any $T$, in any nonequilibrium state, and for any ordered or disordered molecule or material.



\section{Structure of a hot crystal in spacetime and reciprocal spacetime}
\label{section:structure}
Throughout this work I use the convention
that an overbar on any index or variable denotes negation, e.g., ${\bar{\omega}\equiv-\omega}$ and ${\bbk\equiv-\bk}$.
\subsection{The crystal's structure}
\label{section:setup}
I begin by considering a perfect classical crystal, whose 
primitive lattice vectors are ${\{\ba_1,\ba_2,\ba_3\}}$. 
The crystal's perfection might be a consequence of it being at mechanical equilibrium, 
or of it representing the time-averaged structure of a hot crystal at thermal equilibrium.

I notionally partition the crystal into face-sharing primitive unit cells of volume ${\volume\equiv\abs{\ba_1\cdot(\ba_2\times\ba_3)}}$, 
where $\Omega$ denotes the generic primitive cell,
and I choose the \emph{global} origin to be in one of the cells near the center of the crystal.
This means that each cell in the bulk of the crystal contains exactly one point $\bR$ of the infinite
Bravais lattice ${\bravais\equiv\{R^1\ba_1+R^2\ba_2+R^3\ba_3: \,R^1, R^2, R^3 \in\integer\}}$.
This point will identify it and serve as its \emph{local} origin.
The \emph{specific} cell whose origin is at position ${\bR\in\bravais}$ is 
\begin{align*}
\Omega_\bR\equiv\big\{\bR+r^1\ba_1+r^2\ba_2+r^3\ba_3: \bR&\in\bravais, 
\\
r^1,r^2,r^3&\in\big[-1/2,1/2\big)\big\}, 
\end{align*}
and may also be identified simply as cell $\bR$.
${\bR j}$ will identify
the ${j^\text{th}}$ atom in cell ${\Omega_\bR}$; and ${\bR j \alpha}$ will identify the ${\alpha^{\text{th}}}$ lattice
coordinate of atom ${\bR j}$.

If a large crystal is at mechanical equilibrium, all of the cells deep 
within its bulk are identical and have identical environments, but 
the primitive cells near surfaces will be strained
relative to those in the bulk, and the equilibrium lattice 
coordinates of the atoms in surface cells will be different, in general.
Therefore, let us express the displacement of atom ${\bR j}$ from equilibrium as
${\sum_{\alpha=1}^3 w^{j\alpha}(\bR)u^{\bR j\alpha}\ba_\alpha}$, 
where ${w^{j\alpha}:\bravais\to [0,1]}$ is a function whose value is almost one in the bulk of 
the crystal, almost zero at all points of $\bravais$ that are outside the crystal, 
and only differs significantly from the values one and zero near surfaces.
In the bulk, 
the displacement from equilibrium at time $t$ is almost exactly equal to
${\bu^{\bR j}(t) = \sum_{\alpha=1}^3u^{\bR j\alpha}(t)\,\ba_\alpha}$.

I denote a subset of Bravais lattice points, which only contains local origins of {\em bulk} 
cells, by $\bravaisbulk\subset\bravais$.
For simplicity, I choose ${\bravaisbulk}$ to have the same shape and orientation
as the primitive cell, $\Omega$: it is a parallelpiped 
with edges ${(2\nlin+1)\ba_1}$, ${(2\nlin+1)\ba_2}$, and ${(2\nlin+1)\ba_3}$, and
with the origin at its center, such that there are ${\nlin}$ other cells between the origin
cell, $\Omega_{\bzero}$, and the boundary of ${\bravaisbulk}$
in the directions of each of the six vectors ${\pm\ba_1}$, ${\pm\ba_2}$, and ${\pm\ba_3}$.
I choose $\nlin$ to be the largest positive integer for which
the differences between the structures of cells in $\bravaisbulk$, at
both mechanical equilibrium and when the structures are averaged over time,
are negligible. 
Throughout this work, any sum ${\sum_\bR}$ over an unspecified set of lattice vectors 
denotes a sum over all lattice vectors in $\bravaisbulk$.

Although there may be bulk-like cells in ${\bravais\setminus\bravaisbulk}$, from
this point forward the term {\em bulk} refers to the chunk of the crystal's interior, 
${\bulk\equiv\bigcup_{\bR\in\bravaisbulk} \Omega_\bR}$, 
which comprises ${N_c\equiv(2m+1)^3}$ primitive cells, 
and I will restrict my attention to the bulk.
Therefore, for simplicity, I will ignore the set of functions ${\{w^{j\alpha}\}}$ 
and focus on the functions ${\{u^{\bR j\alpha}\}}$; 
I will sometimes refer to the set of all bulk cells simply as \emph{the crystal};
and I will sometimes refer to their boundary with non-bulk cells simply as \emph{the surface}.
However, it is important that we carry with us
the understanding that the
structures and dynamics of cells near surfaces may
differ markedly from bulk cells, and that to ignore them is
to introduce an uncontrolled approximation.

I will be expressing the deviation of the (bulk of the) crystal from mechanical
or thermal equilibrium as 
a superposition of displacements of atoms along a set of vectors, which I will refer
to as the normal mode eigenvectors, or simply the eigenvectors. However, almost all of
the theory presented applies to other complete sets of linearly-independent vectors, 
such as temperature-renormalized sets of eigenvectors derived within statistical 
perturbation theories like the \emph{self-consistent phonon approximation}~\cite{hooton_1955,werthamer_1970,wallace,tadano_2015,bianco_2017,hellman_2013}.

To find a set of mutually-orthogonal normal mode eigenvectors, such that
in the \templim limit
the free motion along each eigenvector is an oscillation with a well-defined
frequency, 
it is necessary to work with ${\psi^{\bR j\alpha}\equiv\sqrt{m_j}u^{\bR j\alpha}}$, rather than ${u^{\bR j\alpha}}$, 
where ${m_j}$ denotes the mass of the $j^\text{th}$ atom in each primitive cell.
Despite its ${\sqrt{\text{mass}}}$ weighting, I will often refer to ${\psi^{\bR j \alpha}}$
as a displacement and its time derivative, ${\dot{\psi}^{\bR j \alpha}}$, as a velocity.

\subsection{Fourier transforming the crystal's structure}
\label{section:fourier_structure}
${\psi^{\bR j\alpha}(t)}$ can be regarded as the value of a function of ${\bR}$ and $t$, i.e., 
the function
\begin{align*}
\psi^{\square j\alpha}:\bravais\times\realone\to\realone; (\bR,t)\mapsto \psi^{\bR j\alpha}(t), 
\end{align*}
where ${\square}$ is a place-holder for the argument of a function.
However, so that the theory that follows is applicable when the 
continuum limit (${\volume\to 0}$) is taken, let us first
define the continuous function
\begin{align*}
&\psi^{j \alpha}_\eta:\realone^3\times\realone\to\realone;
(\br,t)\mapsto \psi^{j \alpha}_\eta(\br,t),
\end{align*}
where
\begin{align*}
\psi^{j\alpha}_\eta(\br,t) \equiv 
\sum_{\bR\in\bravaisbulk}\psi^{\bR j\alpha}(t)\Delta^{(3)}_\eta(\br-\bR),
\end{align*}
and ${\Delta_\eta^{(3)}(\br)}$ is a smooth function whose ${\eta\to 0}$ limit
is the Dirac delta function, ${\delta^{(3)}(\br)}$.
Therefore, 
\begin{align*}
\psi^{j\alpha}_0(\bR,t)\equiv \lim_{\eta\to 0}\psi^{j\alpha}_\eta(\br,t)\eval_{\br=\bR\in\bravaisbulk}=\psi^{\bR j \alpha}(t).
\end{align*}
Then, if ${\tpsi^{j\alpha}(\bk,\omega)}$ denotes the ${\eta\to 0}$ limit of the unitary
Fourier transform ($\fourier$) of ${\psi^{j\alpha}_\eta(\br,t)}$
with respect to both space and time, 
${\psi^{\bR j \alpha}(t)}$ can be expressed as
the superposition of lattice waves, 
\begin{align}
&\psi^{\bR j\alpha}(t)
= 
\sigma^4
\int_{\realone}\dd{\omega} 
\int_{\realone^3}\dd[3]{k}
\tpsi^{j\alpha}(\bk,\omega)
e^{i(\bk\cdot\bR - \omega t)}
\nonumber
\\
& 
=
\sigma^4
\int_{\realone}\dd{\omega} 
\int_{\bzsub}\dd[3]{k}
\sum_\bG
\tpsi^{j\alpha}(\bk+\bG,\omega)
e^{i(\bk\cdot\bR-\omega t)},
\label{eqn:psirja}
\end{align}
where ${\sigma\equiv 1/\sqrt{2\pi}}$. Note that
${\tpsi^{j\alpha *}(\bk,\omega) = \tpsi^{j\alpha}(-\bk,-\omega)}$
follows from the fact that ${\psi^{\bR j \alpha}\in\realone}$.

In Eq.~\ref{eqn:psirja}, each wavevector
has been expressed as the sum, ${\bk+\bG}$, of
a wavevector in the first Brillouin zone, 
${\bk = 2\pi\left(k_1\bb^1 + k_2\bb^2+k_3\bb^3\right)}$, 
and a reciprocal lattice vector, 
${\bG = 2\pi\left(G_1\bb^1 + G_2\bb^2 + G_3\bb^3\right)}$, 
where ${\bb^1}$, ${\bb^2}$, ${\bb^3}$
are the primitive vectors of the reciprocal lattice, ${\bb^\alpha\cdot\ba_\beta=\tensor{\delta}{^\alpha_\beta}}$, 
${k_1,k_2,k_3\in (-0.5,0.5]}$,  and ${G_1, G_2, G_3 \in\integer}$.

\subsubsection{Imposing boundary conditions}
\label{section:boundary_conditions}
Now let us define
\begin{align*}
\tpsi_\bk^{j\alpha}(\omega)\equiv \sigma^3 \dd[3]{k}\sum_\bG\tpsi^{j\alpha}(\bk+\bG,\omega)
=\tpsi_{\bbk}^{j\alpha *}(-\omega),
\end{align*}
and let us use the `surface' boundary conditions (the conditions
at the boundary of the designated bulk),
and the relation 
${e^{i\bk\cdot\bR}=e^{2\pi ik_1R^1}e^{2\pi ik_2R^2}e^{2\pi ik_3R^3}}$,
to  constrain
the set of values of $\bk$ at which  ${\tpsi_{\bk}^{j\alpha}}$
does not necessarily vanish to a finite countable set ${\bza\subset\bz}$.
For example, 
it is straightforward to show that both
the closed boundary conditions,
\begin{align}
\psi^{\pm\nlin\ba_1\,j\alpha}=\psi^{\pm\nlin\ba_2\,j\alpha}=\psi^{\pm\nlin\ba_3\,j\alpha}=0,
\label{eqn:closedbc}
\end{align}
and the 
open boundary conditions
\begin{align}
\partial_\bR \psi^{\bR j\alpha}\eval_{\bR=\pm\nlin\ba_1}
&=\partial_\bR \psi^{\bR j\alpha}\eval_{\bR=\pm\nlin\ba_2}
\nonumber
\\
&=\partial_\bR \psi^{\bR j\alpha}\eval_{\bR=\pm\nlin\ba_3} = 0,
\label{eqn:openbc}
\end{align}
restrict the set $\bza$ to wavevectors $\bk$
for which ${k_1}$, ${k_2}$, and ${k_3}$
are integer multiples of ${(4m)^{-1}}$.
On the other hand, the periodic {\em Born-von K\'arm\'an} boundary conditions,
\begin{align}
\psi^{\bR\pm(2m+1)\ba_1\,j\alpha}&=\psi^{\bR\pm(2m+1)\ba_2\,j\alpha} 
\nonumber \\
&= \psi^{\bR\pm(2m+1)\ba_3\,j\alpha} = \psi^{\bR j \alpha},
\label{eqn:periodicbc}
\end{align}
restrict $\bza$ to wavevectors
$\bk$ for which ${k_1}$, ${k_2}$, and ${k_3}$ are integer multiples of ${(2m)^{-1}}$.

The instantaneous structure of the crystal can be specified by the values of  
three lattice coordinates (${\alpha}$) for each of the $N$ atoms ($j$) in each of the crystal's
$N_c$ primitive cells ($\bR$).  The fact that there are ${3 N N_c}$ degrees of freedom
means that any complete linearly-independent basis capable of specifying
an arbitrary structure of the crystal contains exactly ${3 N N_c}$ elements.
Therefore, since each lattice coordinate ${\psi^{\bR j \alpha}}$ can be
expressed in terms of the elements of the set ${\{\tpsi_\bk^{j\alpha}\}_{\bk\in\bzasub}}$, the number
of elements of ${\bza}$ that are wavevectors of \emph{linearly-independent}
lattice wave contributions to ${\psi^{\bR j \alpha}}$ must be ${N_c}$ - the number
of bulk cells.

After forming set $\bza$, by imposing the constraints on $\bz$ implied by the
boundary conditions, we have
\begin{align}
\psi^{\bR j \alpha} (t)
= \sigma \sum_{\bk\in\bzasub}e^{i\bk\cdot\bR}\int_\realone\dd{\omega}\tpsi^{j\alpha}_\bk(\omega) e^{-i\omega t}.
\label{eqn:bc}
\end{align}
For any set of boundary conditions, time-reversal symmetry implies
that ${-\bk\in\bza}$ if and only if ${\bk\in\bza}$. 
Therefore,  with ${\sum_{\{\bk,\bbk\}}}$ denoting
the sum over all \emph{distinct} pairs ${\{\bk,-\bk\}}$ for which 
${\bk\in\bza}$, and
using ${\psi^{\bR j \alpha *} = \psi^{\bR j \alpha}}$ and 
${\tpsi^{j\alpha *}_\bk(\omega)=\tpsi^{j\alpha}_{\bbk}(-\omega)}$, 
Eq.~\ref{eqn:bc} can be expressed as
\begin{widetext}
\begin{align*}
\psi^{\bR j \alpha}(t)
&= 
\sigma\sum_{\{\bk,\bbk\}}
\int_{\realpos}\dd{\omega}\left[
\tpsi^{j\alpha}_\bk(\omega) e^{i(\bk\cdot\bR-\omega t)}
+
\tpsi^{j\alpha}_\bk(-\omega) e^{i(\bk\cdot\bR+\omega t)}
+
\tpsi^{j\alpha}_{\bbk}(\omega) e^{-i(\bk\cdot\bR+\omega t)}
+
\tpsi^{j\alpha}_{\bbk}(-\omega) e^{-i(\bk\cdot\bR-\omega t)}
\right]
\\
= \psi^{\bR j \alpha *}(t)
&= 
\sigma\sum_{\{\bk,\bbk\}}
\int_{\realpos}\dd{\omega}\left[
\left(\tpsi^{j\alpha}_\bk(\omega)e^{i\bk\cdot\bR} + \tpsi^{j\alpha}_{\bbk}(\omega)e^{-i\bk\cdot\bR}\right)e^{-i\omega t}
+
\left(\tpsi^{j\alpha *}_{\bbk}(\omega)e^{i\bk\cdot\bR} + \tpsi^{j\alpha *}_{\bk}(\omega)e^{-i\bk\cdot\bR}\right)e^{i\omega t}
\right]
\\
\implies
\dpsi^{\bR j \alpha}(t)
& = 
-i\sigma\sum_{\{\bk,\bbk\}}
\int_{\realpos}\dd{\omega}\omega\left[
\left(\tpsi^{j\alpha}_\bk(\omega)e^{i\bk\cdot\bR} + \tpsi^{j\alpha}_{\bbk}(\omega)e^{-i\bk\cdot\bR}\right)e^{-i\omega t}
-
\left(\tpsi^{j\alpha *}_{\bbk}(\omega)e^{i\bk\cdot\bR} + \tpsi^{j\alpha *}_{\bk}(\omega)e^{-i\bk\cdot\bR}\right)e^{i\omega t}
\right].
\end{align*}
\end{widetext}
Now, if the boundary conditions imply that ${\psi^{\bR j \alpha}}$ 
is independent of $t$ at one or more bulk lattice vectors, ${\bR}$, then ${\dpsi^{\bR j \alpha}}$
must vanish at those lattice vectors at all times $t$.
This implies that the coefficients of ${e^{-i\omega t}}$
and ${e^{i\omega t}}$ in the integrand must both
vanish, which can only be
true at all times if ${\tpsi_\bk^{j\alpha *}(\omega) = \tpsi_{\bbk}^{j\alpha}(\omega)}$.
It is straightforward to show that ${\tpsi_\bk^{j\alpha *}(\omega)}$ and ${\tpsi_{\bbk}^{j\alpha}(\omega)}$ would
also be equal if ${\dpsi^{\bR j \alpha}}$ or a higher-order
time derivative of ${\psi^{\bR j \alpha}}$ was independent of time. 

If ${\tpsi^{j\alpha}_\bk(\omega)}$ is expressed in the 
polar form
\begin{align*}
\tpsi^{j\alpha}_\bk(\omega)=\abs{\tpsi^{j\alpha}_\bk(\omega)}\exp\left(i\phi^{j\alpha}_{\bk\omega}\right),
\end{align*}
where 
${\phi^{j\alpha}_{\bk\omega}}$ is a constant,  
then ${\tpsi_\bk^{j\alpha *}(\omega) = \tpsi_{\bbk}^{j\alpha}(\omega)}$
implies that, under either the closed boundary conditions of Eq.~\ref{eqn:closedbc}
or the open boundary conditions of Eq.~\ref{eqn:openbc},
${\phi^{j\alpha}_{\bbk\omega} = -\phi^{j\alpha}_{\bk\omega}}$ and
${\abs{\tpsi^{j\alpha}_{\bbk}(\omega)}=\abs{\tpsi^{j\alpha}_\bk(\omega)}}$.
Therefore, Eq.~\ref{eqn:bc} can be expressed as
\begin{widetext}
\begin{align}
\psi^{\bR j \alpha}(t)
= 4\sigma\sum_{\{\bk,\bbk\}}
\int_{\realpos}\!\dd{\omega}
\abs{\tpsi_\bk^{j\alpha}(\omega)}\cos(\bk\cdot\bR+\phi^{j\alpha}_{\bk\omega})\cos\omega t
= 4\sigma\sum_{\{\bk,\bbk\}}
\int_{\realpos}\!\dd{\omega}
\abs{\tpsi_\bk^{j\alpha}(\omega)}\cos\left(\bk\cdot\bR\right)
\cos(\omega t +\phi^{j\alpha}_{\bk\omega}),
\label{eqn:bc2}
\end{align}
\end{widetext}
where, to reach this expression, we have either assumed that ${\tpsi_\bzero^{j\alpha}(\omega) = 0}$, 
for all ${\omega\in\realpos}$, or we have redefined ${\tpsi_\bzero^{j\alpha}}$ to correct for
the factor of two that arose when ${\sum_{\bk\in\bzasub}}$ was replaced by
${\sum_{\{\bk,\bbk\}}}$. 
Equation~\ref{eqn:bc2} implies that the wave with wavevector ${\bk}$ and frequency ${\omega}$ is a standing wave, 
and that it is the same standing wave as the one with wavevector ${-\bk}$ and frequency ${\omega}$.

Under periodic boundary conditions, it is not true that 
${\tpsi_{\bk}^{j\alpha *}=\tpsi_{\bbk}^{j\alpha}}$, in general. Therefore,
\begin{widetext}
\begin{align}
\psi^{\bR j \alpha} 
&= 2\sigma \sum_{\{\bk,\bbk\}\subset\bzasub}\int_{\realpos}\dd{\omega}
\left[
\abs{\tpsi_{\bk}^{j\alpha}(\omega)}\cos\left(\bk\cdot\bR-\omega t +\phi^{j\alpha}_{\bk\omega}\right)
+
\abs{\tpsi_{\bbk}^{j\alpha}(\omega)}\cos\left(\bk\cdot\bR+\omega t +\phi^{j\alpha}_{\bbk\omega}\right)
\right],
\end{align}
\end{widetext}
which means that, for any given positive frequency $\omega$, 
the contributions to ${\psi^{\bR j \alpha}}$ from any wavevector ${\bk}$ and
its negative, ${-\bk}$, are counter-propagating travelling waves whose amplitudes 
and phases differ, in general.

Note that the term ${\bk=\bzero}$ is independent of ${\bR}$ and can be expressed
as ${2\sigma\abs{\tpsi_{\bzero}^{j\alpha}(\omega)}\cos\left(\omega t + \phi_{\bzero\omega}^{j\alpha}\right)}$.
This term describes a sinusoidally-varying relative displacement of the sublattice of atoms $j$ along
primitive lattice vector ${\ba_\alpha}$, and it can play important 
roles during phase transitions of the crystal.

From now on I will assume that the crystal is finite and bounded. Therefore, 
since not all elements of $\bza$ identify different contributions to $\psi^{\bR j \alpha}$,
$\bzamin$ will denote the subset of ${\bza}$ containing only one wavevector from each pair ${\{\bk,-\bk\}\subset\bza}$.
The number of elements of $\bzamin$ is $N_c$ and the number of elements of ${\bza}$ is ${2N_c-1}$, rather than ${2N_c}$, because
${\bk=0}$  only appears once in $\bza$.

Although we will not be using periodic boundary conditions, and it is not strictly necessary, it will 
often be useful to keep the complex exponential representation of waves and to use set $\bza$
rather than set ${\bzamin}$.
In other words, I will sometimes describe the crystal's structure using more wavevectors than are necessary,
by including terms referenced to the negatives of wavevectors in ${\bzamin}$ in its description.
To facilitate this, let 
${\bzam\equiv\{-\bk:\bk\in\bzamin, \abs{\bk}>0\}}$
and let ${\bzapm\equiv\bzamin\cup\bzam}$.
From now on, ${\sum_{\bk}}$ will denote a sum over all ${\bk\in \bzapm}$, but with
the ${\bk=0}$ term appearing \emph{twice} in the sum.
Therefore, there are ${2 N_c}$ 
terms in ${\sum_{\bk}}$. The sum
${\sum_{\{\bk,\bbk\}}}$ is equivalent to a sum over all ${\bk\in\bzamin}$.

Finally, note that when the integral over $\bza$ was discretized, 
the expression for ${\tpsi_\bk^{j\alpha}(\omega)}$ in terms
of ${\psi^\bR(t)}$ acquired a factor of ${N_c^{-1}}$ and lost a factor of ${\sigma^3}$, to become
\begin{align}
\tpsi_\bk^{j\alpha}(\omega) = \frac{\sigma}{N_c}\sum_\bR\int_\realone\dd{t}
\psi^{\bR j\alpha}(t) e^{-i(\bk\cdot\bR-\omega t)}.
\end{align}
In Sec.~\ref{section:finite_resolution}, when we 
perform an analogous discretization of the integral over time, 
the origin of this factor will become clear.

\subsection{Vector notation}
Appendix~\ref{section:vectors} contains a
discussion of vectors, metrics, metric duals of vectors, and products of vectors, as well as
a more detailed explanation of the vector notation used in this work.
Here I briefly describe this notation.


Boldface type (e.g., $\bu$) is used to denote a vector in $\realone^3$ or $\complex^3$; 
 $\ket{u}$ denotes a vector in ${\realone^{3N}}$ or ${\complex^{3N}}$;
and ${\kket{u}}$ denotes a vector in ${\realone^{3NN_c}}$ or ${\complex^{3NN_c}}$.
The metric duals of the vectors ${\bu}$, ${\ket{u}}$, 
and ${\kket{u}}$ are denoted by ${\bu^\dagger}$, ${\bra{u}}$, and ${\bbra{u}}$,
respectively. 


In ${\complex^{3N}}$ the dual of ${\ket{u}}$ is the operator
${\bra{u}:\complex^{3N}\to\complex; \ket{v}\mapsto \mel{u}{}{v}\equiv \braket{u}{v}}$;
and in ${\complex^{3NN_c}}$ we have
${\bbra{u}:\complex^{3NN_c}\to\complex; \kket{v}\mapsto \bbra{u}\kket{v}\equiv \bbraket{u}{v}}$.
I will sometimes use $\dagger$ to denote the dualizing operation in higher dimensions;
for example ${\ket{u}^\dagger\equiv \bra{u}}$. 

The inner product of two vectors in ${\realone^3}$ is denoted by
${\bu\cdot\bv\equiv g(\bu,\bv)}$, where $g$ denotes the Euclidean metric. The inner product in
${\complex^3}$ is ${\eta(\bu,\bv)\equiv \Re\{\bu\cdot\bv\}=\frac{1}{2}\left[\bu^*\cdot\bv+\bv^*\cdot\bu\right]}$.
In spaces of dimensions ${3N}$ and ${3NN_c}$ 
the inner products are denoted by ${\eta(\ket{u},\ket{v})}$ and ${\eta(\kket{u},\kket{v})}$, respectively, 
where, for example, ${\eta(\kket{u},\kket{v})\equiv \Re\{\bbraket{u}{v}\} = \frac{1}{2}\left[\bbraket{u}{v}+\bbraket{v}{u}\right]}$.
As discussed in Appendix~\ref{section:vectors}, ${\bbraket{u}{v}}$ denotes 
a hybrid of an inner product and a {\em Clifford product}~\cite{hestenes,doran}.

\subsubsection{Vectors specifying the crystal's structure} 
In every vector space, $g_{\alpha\beta}$ will denote ${\ba_\alpha\cdot\ba_\beta\equiv g(\ba_\alpha,\ba_\beta)}$, 
where ${\ba_\alpha}$ and ${\ba_\beta}$ are lattice vectors.

Let ${\bpsi^{\bR j}(t)\equiv \sum_\alpha\psi^{\bR j\alpha}(t)\,\ba_\alpha}$  denote
the $\sqrt{\text{mass}}$-weighted displacement
of atom ${\bR j}$ from equilibrium at time $t$
and let the state of cell $\Omega_\bR$ at time $t$ be specified by
\begin{align*}
\ket{\psi^\bR(t)}\equiv \sum_{j\alpha} \psi^{\bR j\alpha}(t)\ket{j \alpha},
\end{align*}
where  the inner product of the (real) basis vectors %
${\ket{j\alpha}}$ and  ${\ket{i\beta}}$ is 
${\braket{j\alpha}{i\beta}=\delta_{ij}\,\delta_{\alpha\beta}}$.

The state, or microstructure, of the entire crystal at time $t$ can be specified by
\begin{align*}
\kket{\psi(t)}\equiv \sum_{\bR j \alpha} \psi^{\bR j\alpha}(t)\kket{\bR j \alpha}\in\realone^{3 N N_c}, 
\end{align*}
where ${\bbraket{\bR j\alpha}{\bR' i\beta}=\delta_{\bR\bR'}\,\delta_{ij}\,\delta_{\alpha\beta}}$.

I will denote the vectors in ${\complex^3}$ and ${\complex^{3N}}$ whose
components are the Fourier transforms, $\tpsi_{\bk}^{j\alpha}(\omega)$, of the components, 
${\psi^{\bR j\alpha}(t)}$, 
of ${\bpsi^{\bR j}(t)}$  and
${\ket{\psi^\bR(t)}}$, 
by
${\btpsi_\bk^j(\omega)}$ and
${\ket{\tpsi_\bk(\omega)}}$, respectively.
Therefore Eq.~\ref{eqn:bc} can also be expressed in the following forms:
\begin{align}
\ket{\psi^\bR(t)} &= 
\sigma
\sum_\bk 
\int_{\realone}\dd{\omega}
\ket{\tpsi_\bk(\omega)}
e^{i(\bk\cdot\bR-\omega t)}
\label{eqn:ft1}
\\
&= 
\sigma
\sum_\bk e^{i\bk\cdot\bR}\int_{\realpos}\dd{\omega}
\bigg[
e^{-i\omega t}
\ket{\tpsi_\bk(\omega)}
\nonumber
\\
&\qquad\qquad\qquad\qquad\quad +
e^{i\omega t}
\ket{\tpsi_{\bbk}^*(\omega)}
\bigg].
\label{eqn:ft2}
\end{align}

\subsection{Eigenvectors, cell eigenvectors, and polarization vectors}
\label{section:eigenbasis}
In Sec.~\ref{section:normal_modes}, as is traditional, I will
begin the development of phonon theory 
by Taylor expanding the potential energy in atomic displacements from equilibrium; retaining
only the lowest-order non-vanishing and non-trivial terms, which
are those at second order; and using the second derivatives
with respect to atomic positions to define
a matrix whose eigenvectors are
the crystal's \emph{normal mode eigenvectors}. 

However, almost all of the theory that follows, except that which is 
presented in Secs.~\ref{section:normal_modes} and~\ref{section:hermitian}, can 
be derived without any discussion of energetics.
It applies equally if the set of ${3 N N_c}$ 
eigenvectors, 
${\{\kket{E_\mu}\}_{\mu=1}^{3NN_c}}$, is not the set of
normal mode eigenvectors, but 
a complete orthonormal basis of ${\realone^{3NN_c}}$, which is arbitrary apart
from the requirement that its lattice components respect the periodicity
of the crystal's bulk, in the sense described at the beginning of Sec.~\ref{section:eigenvector_symmetry}.

For familiarity, I will use the term \emph{eigenvector} to refer to an element of such a basis. 
Doing so is not a mathematical crime because each element
is an eigenvector of any operator of the form ${\sum_{\mu=1}^{3NN_c}\xi_\mu\kket{E_\mu}\bbra{E_\mu}}$, 
for any set of ${3NN_c}$ scalars, $\xi_\mu$.
I will also use language that suggests that the eigenvectors are normal mode eigenvectors 
and that the corresponding eigenvalues are the normal mode frequencies. However, it is
to be understood that the theory has greater generality.

Pointing out its generality is important, because it helps to clarify which aspects
of phonon theory are determined by energetics, and which are mathematical artefacts
of a crystal's structure and symmetry.

\subsubsection{Normal mode eigenvectors}
\label{section:normal_modes}
For a given set of initial conditions, 
the dynamics of the crystal in the \templim limit are defined
by the set 
of normal mode eigenvectors and their
frequencies ${\{\omega_\mu\}_{\mu=1}^{3NN_c}}$. 
Initial conditions determine the phases of the lattice
waves at ${t=0}$ and their amplitudes.
I will denote the set of projections of the normal mode eigenvectors
onto the ${3NN_c}$-dimensional vector subspace whose elements specify structures 
of the crystal's {\em bulk} by ${\{\kket{E_\mu}\}_{\mu=1}^{3NN_c}}$ .

By doing this, I am cheating: There are more than ${3NN_c}$ normal modes,
because the crystal has more than ${3NN_c}$ degrees of freedom when
atoms outside the bulk are counted. 
There are ways to mitigate this problem, 
such as by expressing the vector space spanned by the true eigenvectors
as a direct sum or direct product of a `surface' vector space and
the bulk vector space ${\realone^{3 N N_c}}$, and transforming
the set of eigenvectors into the union of a bulk basis and a surface
basis, in some optimal way. The optimal transformation will depend
on the boundary conditions and I do not discuss it further
because I have not explored it further. However, 
it is important to remember that these gaps exist in the logic of this work, and 
in the mathematical infrastructure assembled within it. 

At small finite $T$ it can be useful to approximate the crystal's dynamics as 
a superposition of harmonic oscillations along the normal mode 
eigenvectors, but this approximation becomes less accurate as $T$ increases.
However, because
the set of eigenvectors is a complete basis of the crystal's ${3NN_c}$-dimensional configuration space, it is 
always mathematically possible to express the instantaneous structure of the crystal, relative
to its structure in the \templim limit, as a superposition of normal mode eigenvectors. 
In this section I show how to use the periodicity of the crystal's bulk to express
the structure of each bulk primitive cell $\Omega_\bR$ in a basis of {\em cell eigenvectors}, where
each cell eigenvector is parallel to the projection of a different crystal eigenvector
onto the $3N$-dimensional subspace of 
${\realone^{3NN_c}}$ spanned by ${\{\kket{\bR j \alpha}: j\in\{1,\cdots, N\},\;\alpha\in\{1,2,3\}\}}$.

In the low $T$ limit 
the equations of motion of the atoms can be expressed as 
\begin{align}
m_j \,
\ddot{u}^{\bR j\alpha}(t)  
&= -\sum_{\bR' i \beta} 
H^{\bR j \alpha}_{\bR' i \beta}\,
u^{\bR 'i\beta}(t)  
\nonumber\\
\implies\ddot{\psi}^{\bR j\alpha} &= 
-\sum_{\bR' i \beta } D^{\bR j \alpha}_{\bR' i \beta}
\,
\psi^{\bR'i\beta},
\label{eqn:eom}
\end{align}
where ${H^{\bR j \alpha}_{\bR' i \beta} \equiv \pdv*[2]{U}{u^{\bR j \alpha}}{u^{\bR' i \beta}}\eval_{\bu=0}}$ 
is an element of the crystal's lattice-coordinate Hessian matrix evaluated at mechanical equilibrium; $U$ is the potential 
energy; and ${D^{\bR j \alpha}_{\bR' i \beta}\equiv H^{\bR j \alpha}_{\bR' i \beta}/\sqrt{m_i m_j}}$.

Defining ${\Dyn\equiv \sum_{\bR\bR'}\sum_{j\alpha\,i\beta} D^{\bR j \alpha}_{\bR' i \beta} \ddyad{\bR j \alpha}{\bR' i \beta}}$, 
allows Eq.~\ref{eqn:eom} to be expressed as ${\kket{\ddot{\psi}}=-\Dyn\kket{\psi}}$. 
Then, using the fact that
${\kket{\ddot{\psi}}=-\omega_\mu^2\kket{\psi}}$ if and only if
${\kket{\psi}\propto\kket{E_\mu}}$, Eq.~\ref{eqn:eom} leads to the equation ${\Dyn \kket{E_\mu} = \omega_\mu^2\kket{E_\mu}}$, 
whose solutions are the eigenvectors. 
By defining the lattice components ${E_\mu^{\bR j \alpha}\equiv \bbraket{\bR j \alpha}{E_\mu}}$ of eigenvector ${\kket{E_\mu}}$, 
it can be expressed as ${\kket{E_\mu}\equiv \sum_{\bR j \alpha}E_\mu^{\bR j  \alpha}\kket{\bR j \alpha}}$,
and the eigenvalue equation
can also be expressed as
\begin{align}
\sum_{\bR' i \beta} 
D^{\bR j \alpha}_{\bR' i \beta}
E_\mu^{\bR' i \beta}  &= \omega_\mu^2 E_\mu^{\bR j \alpha}.
\label{eqn:eigen}
\end{align}

\subsubsection{Arbitrary `eigenvector' basis}
\label{section:arbitrary_eigenvector}
It will be convenient to express the lattice components of ${\kket{E_\mu}}$ in the Fourier-expanded form 
\begin{align}
E^{\bR j \alpha}_\mu &\equiv \frac{1}{\sqrt{2N_c}}\sum_\bk\eig_{\bk\mu}^{j\alpha} e^{i\bk\cdot\bR} 
\label{eqn:FTeig}
\\
&= 
\frac{1}{\sqrt{2N_c}}\sum_{\{\bk,\bbk\}}\left[
\eig_{\bk\mu}^{j\alpha} e^{i\bk\cdot\bR} 
+
\eig_{\bbk\mu}^{j\alpha} e^{-i\bk\cdot\bR} 
\right]
\label{eqn:FTeig2}
\end{align}
where ${\eig_{\bk\mu}^{j\alpha}\in\complex}$. 
If we multiply both sides of Eq.~\ref{eqn:FTeig} by ${\cos(\bk'\cdot\bR)=\frac{1}{2}(e^{i\bk'\cdot\bR}+e^{-i\bk'\cdot\bR})\in\realone}$ 
and sum over all ${\bR}$ we get
\begin{align*}
&\sum_\bR E_\mu^{\bR j \alpha} \cos\left(\bk'\cdot\bR\right)
\\
&=\frac{1}{2\sqrt{2N_c}}
\sum_{\bk} \eig_{\bk\mu}^{j\alpha} \bigg[\sum_\bR e^{i(\bk+\bk')\cdot\bR} 
+ \sum_\bR e^{i(\bk-\bk')\cdot\bR}\bigg]
\\
&=\frac{1}{2}\sqrt{\frac{N_c}{2}}
\sum_{\bk} \eig_{\bk\mu}^{j\alpha} \left(\delta_{\bk\bbk'}+\delta_{\bk\bk'}\right)
= \frac{1}{2} \sqrt{\frac{N_c}{2}}\left(\eig_{\bk'\mu}^{j\alpha}+\eig_{\bbk'\mu}^{j\alpha}\right),
\end{align*}
and, since ${E_{\mu}^{\bR j \alpha}}$ and ${\cos(\bk'\cdot\bR)}$ are real for all 
$\bR$, ${\eig_{\bk'\mu}^{j\alpha}+\eig_{\bbk'\mu}^{j\alpha}}$ must be real.
Therefore ${\eig_{\bk\mu}^{j\alpha *}=\eig_{\bbk\mu}^{j\alpha}}$, for all ${\bk\in\bzamin}$.
It follows that, if the argument of ${\eig_{\bk\mu}^{j\alpha}}$ is denoted by ${\vartheta_{\bk\mu}^{j\alpha}}$, Eq.~\ref{eqn:FTeig2} can be expressed as 
\begin{align*}
E_{\mu}^{\bR j \alpha}
& = \sqrt{\frac{2}{N_c}}\sum_{\{\bk,\bbk\}}\abs{\eig_{\bk\mu}^{j\alpha}}\cos\left(\bk\cdot\bR + \vartheta_{\bk\mu}^{j\alpha}\right).
\end{align*}

\subsubsection{Eigenvector symmetry}
\label{section:eigenvector_symmetry}
Let ${r_{\mu}^{\bR j\alpha}(\dbR)\equiv E_\mu^{\bR j \alpha}/E_\mu^{\bR+\dbR\,j\alpha}}$
denote the ratio of the eigenvector coordinates of atoms ${\bR j \alpha}$ and ${\bR+\dbR j \alpha}$.
Then, 
\begin{align}
\sum_{\bk}\eig_{\bk\mu}^{j\alpha} e^{i\bk\cdot\bR} 
=
r_{\mu}^{\bR j\alpha}(\dbR)\sum_{\bk} 
\eig_{\bk\mu}^{j\alpha} e^{i\bk\cdot(\bR+\dbR)},
\label{eqn:ratio0}
\end{align}
and, if we assume that ${r_\mu^{\bR j\alpha}(\dbR)}$ is approximately independent of $\bR$
within the crystal's bulk, 
we can multiply both sides by ${e^{-i\bk'\cdot\bR}}$, sum over ${\bR}$, 
and use orthogonality to find
\begin{align*}
\sum_{\bk}\eig_{\bk\mu}^{j\alpha} \left[1-
r_{\mu}^{\bR j\alpha}(\dbR)
e^{i\bk\cdot\dbR}\right]
\left(\delta_{\bk\bk'}  +\delta_{\bk\bbk'}\right) &=0
\\
\Rightarrow
\abs{\eig_{\bk\mu}^{j\alpha}}\left[\cos\vartheta_{\bk\mu}^{j\alpha} - r_\mu^{\bR j\alpha}(\dbR)\cos(\bk\cdot\dbR+\vartheta_{\bk\mu}^{j\alpha})\right] &= 0.
\end{align*}
This means that, for every ${\bk\in\bzamin}$,
either ${\big|\eig_{\bk\mu}^{j\alpha}\big|=0}$,
or ${r_\mu^{\bR j \alpha}(\dbR)= \cos\vartheta_{\bk\mu}^{j\alpha}/\cos(\bk\cdot\dbR+\vartheta_{\bk\mu}^{j\alpha})}$
or both. 
However, ${r_\mu^{\bR j \alpha}}$ has a mode ($\mu$) dependence but
no explicit wavevector dependence. 
Therefore the fact that, when ${\eig_{\bk\mu}^{j\alpha}}$ does not vanish,
${r_\mu^{\bR j \alpha}}$ depends on the wavevector $\bk$ that labels ${\eig_{\bk\mu}^{j\alpha}}$,
implies that only one wavevector pair ${\{\bk,\bbk\}}$ contributes to each eigenvector ${\kket{E_\mu}}$.

From this point forward, each mode will be labelled by a pair of indices, $\bk\mu$, 
where $\mu$ is now an index that runs over all modes whose wavevector pair is ${\{\bk,\bbk\}}$.
For example, the frequency and eigenvector of mode ${\bk\mu}$ will be denoted
by ${\omega_{\bk\mu}}$ and ${\kket{E_{\bk\mu}}\equiv \sum_{\bR j \alpha}E_{\bk\mu}^{\bR j \alpha}\kket{\bR j \alpha}}$, respectively, 
where 
\begin{align}
E_{\bk\mu}^{\bR j \alpha}
&\equiv \frac{1}{\sqrt{2N_c}}\left(\eig_{\bk\mu}^{j\alpha}e^{i\bk\cdot\bR}+ \eig_{\bbk\mu}^{j\alpha}e^{-i\bk\cdot\bR}\right) 
=E_{\bbk\mu}^{\bR j \alpha}.
\label{eqn:eigenvector0}
\end{align}
As expected from the discussion in Sec.~\ref{section:fourier},
the eigenvectors of modes ${\bk\mu}$ and ${\bbk\mu}$ are the same (${\kket{E_{\bk\mu}}=\kket{E_{\bbk\mu}}}$), 
because ${\bk\mu}$ and ${\bbk\mu}$ are different labels for the same mode.
Obviously their eigenvalues are also equal, i.e., ${\omega_{\bk\mu}^2=\omega_{\bbk\mu}^2}$.
Therefore, when working with {\em normal modes} of the crystal, there is no reason to distinguish between
modes ${\bk\mu}$ and ${\bbk\mu}$, because each normal mode is a standing wave comprised
of a counter-propagating pair of travelling waves whose amplitudes are equal.

In the \templim~limit, we do not need to decompose the standing wave into a pair of counterpropagating waves
labelled by ${\bk\mu}$ and ${\bbk\mu}$, but at finite temperature this decomposition is useful
because finite-temperature phonons are not standing waves but travelling wave packets, which have a propagation direction.
Therefore, in Sec.~\ref{section:eigenbasis_expansion}, we will distinguish between them.

I will refer to $\mu$ as the {\em branch index} or {\em band index} 
and I will refer to the quasiconnected set of points ${\{(\bk,\omega_{\bk\mu}):\bk\in\bza\}}$ as
the $\mu^{\text{th}}$ band.
Note that $E_{\bk\mu}^{\bR j  \alpha}$ can also
be expressed in the form
\begin{align}
E_{\bk\mu}^{\bR j  \alpha} = 
\sqrt{\frac{2}{N_c}}\abs{\eig_{\bk\mu}^{j \alpha}} \cos\left(\bk\cdot\bR+\vartheta_{\bk\mu}^{j\alpha}\right) = E_{\bbk\mu}^{\bR j \alpha}
\label{eqn:eigenvector}
\end{align}

The potential energy, $U$, is a smooth real function of atomic positions. Therefore,
${D^{\bR j\alpha}_{\bR' i\beta}=D^{\bR' i\beta}_{\bR j\alpha}}$ for all pairs of components, 
${\bR j \alpha}$ and ${\bR' i \beta}$. It follows that ${\Dyn}$ is Hermitian (symmetric, in fact) and that any two eigenvectors,
${\kket{E_{\bk\mu}}}$ and
${\kket{E_{\bk'\nu}}}$, with different eigenvalues, ${\omega_{\bk\mu}^2}$ and ${\omega_{\bk'\nu}^2\neq\omega_{\bk\mu}^2}$, 
are orthogonal. Multiple eigenvectors can only have precisely the same eigenvalues if they are equivalent by symmetry.
When this is the case, they can be chosen to be orthogonal.
Therefore I choose the set of eigenvectors to be orthonormal.
By this I mean that ${\bbraket{E_{\bk\mu}}{E_{\bk\mu}}=\bbraket{E_{\bk\mu}}{E_{\bbk\mu}}=1}$ and
that if ${\mu\neq \nu}$ or if ${\bk'\notin \{\bk,\bbk\}}$, 
then ${\bbraket{E_{\bk\mu}}{E_{\bk'\nu}}=0}$.
We can express this, as usual, in the form 
\begin{align}
\bbraket{E_{\bk\mu}}{E_{\bk'\nu}}=\sum_{\bR j}\sum_{\alpha\beta} E_{\bk\mu}^{\bR j\alpha *}g_{\alpha\beta} E_{\bk'\nu}^{\bR i\beta}=\delta_{\bk\bk'}\delta_{\mu\nu} .
\label{eqn:Eorthogonality}
\end{align}

For each pair ${\bk\mu}$ that labels a wavevector and a branch, let us define a vector
${\ket{\eig_{\bk\mu}}\equiv\sum_{j\alpha} \eig_{\bk\mu}^{j\alpha}\ket{j\alpha}\in\complex^{3N}}$.
The inner product of two such vectors is
\begin{align*}
\braket{\eig_{\bk\mu}}{\eig_{\bk'\nu}}=\sum_{j\alpha\beta}\eig_{\bk\mu}^{j\alpha *}g_{\alpha\beta}\eig_{\bk'\nu}^{j\beta}
=\braket{\eig_{\bbk'\nu}}{\eig_{\bbk\mu}}.
\end{align*}
Therefore, inserting 
Eq.~\ref{eqn:eigenvector0} into Eq.~\ref{eqn:Eorthogonality} and using orthogonality relations 
gives
\begin{widetext}
\begin{align*}
\bbraket{E_{\bk\mu}}{E_{\bk'\nu}}
 &= \frac{1}{2N_c}
\sum_{\bR}
\left[
\braket{\eig_{\bk\mu}}{\eig_{\bk'\nu}}e^{i(\bk'-\bk)\cdot\bR}
+
\braket{\eig_{\bbk\mu}}{\eig_{\bk'\nu}}e^{i(\bk'+\bk)\cdot\bR}
+
\braket{\eig_{\bk\mu}}{\eig_{\bbk'\nu}}e^{-i(\bk'+\bk)\cdot\bR}
+
\braket{\eig_{\bbk\mu}}{\eig_{\bbk'\nu}}e^{-i(\bk'-\bk)\cdot\bR}
\right],
\\
&
 = \frac{1}{2}
\left[
\braket{\eig_{\bk\mu}}{\eig_{\bk'\nu}}
\delta_{\bk\bk'}
+
\braket{\eig_{\bbk\mu}}{\eig_{\bk'\nu}}
\delta_{\bk\bbk'}
+
\braket{\eig_{\bk\mu}}{\eig_{\bbk'\nu}}
\delta_{\bk\bbk'}
+
\braket{\eig_{\bbk\mu}}{\eig_{\bbk'\nu}}
\delta_{\bk\bk'}
\right] = 
\delta_{\bk\bk'}\delta_{\mu\nu}.
\end{align*}
\end{widetext}
If we use ${\eig_{\bk\mu}^{j\alpha *}=\eig_{\bbk\mu}^{j\alpha} \implies \ket{\eig_{\bk\mu}}^\dagger=\bra{\eig_{\bbk\mu}}}$, 
which leads to ${\braket{\eig_{\bk\mu}}{\eig_{\bbk'\nu}}=\braket{\eig_{\bk'\nu}}{\eig_{\bbk\mu}}}$
and similar expressions, we can simplify this to the form
\begin{align*}
\braket{\eig_{\bk\mu}}{\eig_{\bk'\nu}}\delta_{\bk\bk'} + \braket{\eig_{\bk\mu}}{\eig_{\bbk'\nu}}\delta_{\bk\bbk'}
=
\delta_{\bk\bk'}\delta_{\mu\nu}.
\end{align*}
Now ${\bk=\bk'}$ implies that ${\bk\neq\bbk'}$ and we are left with
\begin{align}
\braket{\eig_{\bk\mu}}{\eig_{\bk\nu}}=\delta_{\mu\nu}.
\label{eqn:ortho}
\end{align}
Equation~\ref{eqn:ortho} means that the set of all vectors ${\ket{\eig_{\bk\mu}}}$ {\em at the same wavevector} $\bk$
is an orthonormal set.

\subsubsection{Constraints of symmetry and the number of DOFs}
Given a mode ${\bk\mu}$, and values of ${j}$ and ${\alpha}$, the function
${E_{\bk\mu}^{\square j \alpha}: \bravaisbulk\to \realone; \bR\mapsto E_{\bk\mu}^{\bR j \alpha}}$ 
assigns a real number to each of the $N_c$ elements of $\bravaisbulk$; and 
the function ${E_{\bk\mu}^{\square i\beta}}$, where ${i\beta\neq j\alpha}$, assigns a different
set of $N_c$ real numbers to ${\bravaisbulk}$. Therefore ${\kket{E_{\bk\mu}}}$ is defined by
the set of ${3N}$ functions,
${\{E_{\bk\mu}^{\square j \alpha}:1\leq j \leq N, \alpha\in\{1,2,3\}\}}$, 
each of which assigns $N_c$ real numbers to $\bravaisbulk$.

Equation~\ref{eqn:eigenvector0} implies that if the function
${\eig_{\square \mu}^{j \alpha}:\bza\to \complex; \bk\mapsto \eig_{\bk\mu}^{j \alpha}}$ 
is known, the function ${E_{\bk\mu}^{\square j \alpha}}$ can be calculated.
Therefore the information possessed by the function ${\eig_{\square\mu}^{j\alpha}}$
must be equivalent to the information possessed by the function ${E_{\bk\mu}^{\square j \alpha}}$, 
which means that ${\eig_{\square\mu}^{j\alpha}}$ must assign exactly $N_c$ real numbers 
to the set ${\bzamin}$.

However
Eq.~\ref{eqn:eigenvector} suggests that, if each ${\eig_{\bk\mu}^{j\alpha}}$
is complex, the function ${\eig_{\square\mu}^{j\alpha}}$ assigns \emph{two} numbers, ${\abs{\eig_{\bk\mu}^{j\alpha}}}$ and
${\vartheta_{\bk\mu}^{j\alpha}}$, for every one of the ${N_c}$ numbers, 
${\{E_{\bk\mu}^{\bR j \alpha}:\bR \in \bravaisbulk\}}$, assigned by
${E_{\bk\mu}^{\square j \alpha}}$.
However, ${\abs{\eig_{\bk\mu}^{j\alpha}}}$ is positive, so this discrepancy
is resolved if ${\vartheta_{\bk\mu}^{j\alpha}\in\{0,\pi\}}$, which implies
that ${\eig_{\bk\mu}^{j\alpha}}$ is real.
Therefore, from 
Eq.~\ref{eqn:eigenvector}, we find that 
\begin{align}
E_{\bk\mu}^{\bR j  \alpha} = 
\pm\sqrt{\frac{2}{N_c}}\abs{\eig_{\bk\mu}^{j \alpha}} \cos\left(\bk\cdot\bR\right) = E_{\bbk\mu}^{\bR j \alpha}.
\label{eqn:eigenvector2}
\end{align}
Although this discussion establishes that each ${\eig_{\bk\mu}^{j\alpha}}$ is real, 
it will be useful in later versions of this work to assume that it is complex.

Since there must be ${3NN_c}$ independent degrees of freedom, or \emph{eigenvectors}, 
and since ${3NN_c}$ real numbers are required to express each one in terms of the displacements from equilibrium, 
the set of all
${E_{\bk\mu}^{\bR j \alpha}}$ must consist of ${(3NN_c)^2}$ real numbers.
Therefore, there must be $3N$ modes ${\bk\mu}$ for each of the ${N_c}$ 
elements $\bk$ of $\bza$.

\subsubsection{Dynamical matrix}
\label{section:hermitian}
Another way to see that there are
exactly ${3N}$ modes whose wavevector is ${\bk}$, which
means that the total number of modes is ${3NN_c}$, 
is to take the more traditional energetic route.
This involves showing 
that when 
${\{\kket{E_{\bk\mu}}\}}$ is the set of normal mode eigenvectors, 
${\{\ket{\eig_{\bk\mu}}\}_{\mu=1}^{3N}}$ is 
the set of eigenvectors of a Hermitian operator on a ${3N}$-dimensional vector space. This implies that it is either an orthogonal set
or can be made orthogonal, which makes the proof of Eq.~\ref{eqn:ortho} redundant. 

When studying vibrations in the bulk of the crystal, the 
dimensionality of the problem can be reduced from ${3NN_c}$ to ${3N}$ by
expressing Eq.~\ref{eqn:eigen} as a set of $N_c$ equations, 
with one equation for each primitive cell $\bR$.
Then, because all bulk cells are identical in the \templim limit,
or when averaged over time, for most purposes 
it is only necessary to study a single representative bulk cell.
To achieve this simplification, Eq.~\ref{eqn:eigenvector0}
can be inserted into Eq.~\ref{eqn:eigen}, with ${\bR'\equiv\bR+\dbR}$, to give
\begin{align}
\sum_{i \beta} 
&\sum_{\dbR} 
D^{\bR j \alpha}_{\bR+\dbR i \beta} 
\left(
\eig^{i\beta}_{\bk\mu} 
e^{i\bk\cdot(\bR+\dbR)}
+
\eig^{i\beta}_{\bbk\mu} 
e^{-i\bk\cdot(\bR+\dbR)}
\right)
\nonumber
\\
&=\sum_{i\beta}\mel{j\alpha}{\tDyn_\bk(\bR)}{i\beta} 
\left(
\eig_{\bk\mu}^{i\beta} e^{i\bk\cdot\bR}
+
\eig_{\bbk\mu}^{i\beta} e^{-i\bk\cdot\bR}
\right)
\nonumber
\\
&= \omega_{\bk\mu}^2  
\left(
\eig_{\bk\mu}^{i\beta} e^{i\bk\cdot\bR}
+
\eig_{\bbk\mu}^{i\beta} e^{-i\bk\cdot\bR}
\right)
\label{eqn:eigen2}
\end{align}
where simplification has been achieved by replacing the sum over ${\dbR}$ of one
of the exponential terms on the first line by a sum over ${-\dbR}$, and
where the Hermitian operator ${\tDyn_\bk(\bR):\realone^{3N}\to\realone^{3N}}$ is defined
as the operator whose matrix elements are
\begin{align*}
\mel{j\alpha}{\tDyn_\bk(\bR)}{i\beta}
\equiv \sum_{\dbR} \bigg(
D^{\bR j \alpha}_{\bR+\dbR i \beta}
+
D^{\bR j \alpha}_{\bR-\dbR i \beta}
\bigg)
e^{i\bk\cdot\dbR}.
\end{align*}
Using the orthogonality of ${e^{i\bk\cdot\bR}}$ and ${e^{-i\bk\cdot\bR}}$ as
functions of $\bR$, Equation~\ref{eqn:eigen2} can be recast into the form
\begin{align*}
\tDyn_\bk(\bR)\ket{\eig_{\bk\mu}}=\omega^2_{\bk\mu}\ket{\eig_{\bk\mu}},
\end{align*}
which has exactly ${3N}$ solutions ${\ket{\eig_{\bk\mu}}}$ of unit norm that differ from one another 
by more than phase factor. Therefore, ${\ket{\eig_{\bk\mu}}}$
and ${\ket{\eig_{\bbk\mu}}}$ must differ only by a phase, which implies that ${\vartheta^{j\alpha}_{\bk\mu}}$
is independent of both ${j}$ and ${\alpha}$ and we will denote it by ${\vartheta_{\bk\mu}=-\vartheta_{\bbk\mu}}$.
Having already assumed that all bulk cells are equivalent, we
can ignore the dependences of ${D^{\bR j\alpha}_{\bR+\dbR\,i\beta}}$
and
${\tDyn_\bk(\bR)}$ on $\bR$ and regard each solution ${\ket{\eig_{\bk\mu}}}$
as independent of $\bR$ in the bulk. 

\subsubsection{Cell eigenvectors}
There is a different set of ${3N}$ solutions 
for each pair ${\{\bk,\bbk\}\subset\bza}$. 
Now, because the phase of each component ${\eig_{\bk\mu}^{j\alpha}}$
of ${\ket{\eig_{\bk\mu}}}$ is the same, it is simple to define a real vector
\begin{align*}
\ket{\neig_{\bk\mu}}\equiv\sum_{j\alpha}\neig_{\bk\mu}^{j\alpha}\ket{j\alpha}\equiv \frac{1}{\sqrt{2}}\left[\ket{\eig_{\bk\mu}}+\ket{\eig_{\bbk\mu}}\right]\in\realone^{3N},
\end{align*}
which can easily be shown to 
satisfy the conditions 
\begin{align*}
\ket{\neig_{\bk\mu}}&=\ket{\neig_{\bbk\mu}}\\
\braket{\neig_{\bk\mu}}{\neig_{\bk\nu}} &= \delta_{\mu\nu}\\
\tDyn_\bk(\bR)\ket{\neig_{\bk\mu}}&=\omega^2_{\bk\mu}\ket{\neig_{\bk\mu}}.
\end{align*}
Each vector $\ket{\neig_{\bk\mu}}$ 
specifies how atoms in a single primitive cell of the bulk move
when only mode ${\bk\mu}$ is active.  
The vectors ${\{\ket{\neig_{\bk\mu}}\}_{\mu=1}^{3N}}$ 
are known as the {\em (primitive) cell eigenvectors}.
It is important to note that, in general, and depending on the boundary conditions, 
their coordinates, ${\neig_{\bk\mu}^{j\alpha}}$, cannot be used in place of ${\eig_{\bk\mu}^{j\alpha}}$ in
Eq.~\ref{eqn:eigenvector0}.

It is often useful to know each mode's pattern of displacements, rather
than its pattern of weighted displacements. Therefore, let us define 
the real symmetric operator
\begin{align*}
\Mss\equiv\sum_{j\alpha\,i\beta} \sqrt{m_i}\delta_{ij}\delta_{\alpha\beta}\dyad{j\alpha}{i\beta}=\sum_{j\alpha}\sqrt{m_j}\dyad{j\alpha},
\end{align*}
and denote its inverse by
${\Mss^{-1}}$. The vector 
\begin{align*}
\ket{\pol_{\bk\mu}}\equiv \sum_{\bk\mu}\pol_{\bk\mu}^{j\alpha}\ket{j\alpha}\equiv\Mss^{-1}\ket{\neig_{\bk\mu}}\in\realone^{3N}, 
\end{align*}
which is known as the {\em polarization vector} of 
mode ${\bk\mu}$, specifies the pattern of unweighted displacements in the bulk of the crystal.

The set of eigenvalue equations can be used to deduce a set of equations
whose solutions are the polarization vectors as follows,
\begin{align}
\tDyn_\bk(\bR)\ket{\neig_{\bk\mu}} & = \omega_{\bk\mu}^2 \ket{\neig_{\bk\mu}}  
\nonumber \\
\implies
\tDyn_\bk(\bR) \Mss\Mss^{-1} \ket{\neig_{\bk\mu}}&=\omega_{\bk\mu}^2 \Mss\Mss^{-1} \ket{\neig_{\bk\mu}}
\nonumber \\
\therefore
\tHess_\bk(\bR)\ket{\pol_{\bk\mu}}
&=\omega_{\bk\mu}^2\ket{\pol_{\bk\mu}}, 
\end{align}
where ${\tHess_\bk(\bR)\equiv \Mss^{-1}\tDyn_\bk(\bR)\Mss}$.
Note that, in general, ${\tHess_\bk(\bR)}$ is not symmetric or Hermitian
and the polarization vectors are not mutually orthogonal. The
orthonormality of the set ${\{\ket{\neig_{\bk\mu}}\}_{\mu=1}^{3N}}$
leads directly to the generalized orthogonality relation
\begin{align}
\braket{\Mss\pol_{\bk\mu}}{\Mss\pol_{\bk\nu}} 
= \sum_{j \alpha \beta} m_j \pol_{\bk\mu}^{j\alpha *}g_{\alpha\beta}\pol_{\bk\nu}^{j\beta}=\delta_{\mu\nu}.
\label{eqn:polortho}
\end{align}
When the masses of all atoms are equal, the polarization vectors 
are mutually orthogonal and parallel to the cell eigenvectors.

Note that the normal mode eigenvectors ${\kket{E_{\bk\mu}}}$ are real, 
as are all of the other quantities mentioned in Sec.~\ref{section:eigenbasis}, 
except complex exponentials and the vectors ${\ket{\eig_{\bk\mu}}}$. However, 
we did not really need to introduce either those vectors or the complex exponentials in this 
section. We could have worked with real trigonometric functions instead.
Therefore, so far, we have not really needed to work in complex vector spaces.

However, starting in Sec.~\ref{section:complex_waves}, it will become 
useful to work with complex vectors.
Therefore, from this point forward each vector in ${\realone^3}$, ${\realone^{3N}}$, or ${\realone^{3 N N_c}}$, 
should be regarded as a real-valued element of
complex vector space 
${\complex^3\supset\realone^3}$, 
${\complex^{3N}\supset\realone^{3N}}$, 
or
${\complex^{3 N N_c}\supset\realone^{3 N N_c}}$, respectively.
For example, ${\ket{\psi^\bR}\in \complex^{3N}}$, ${\kket{E_{\bk\mu}}\in\complex^{3 N N_c}}$ and ${\kket{\bR j \alpha}\in\complex^{3 N N_c}}$.
The domain and codomain of ${\Dyn}$ are also extended from ${\realone^{3 N N_c}}$ to ${\complex^{3 N N_c}}$.
The basis sets ${\{\ba_\alpha\}}$, ${\{\ket{j\alpha}\}}$, and ${\{\kket{\bR j \alpha}\}}$ of 
${\realone^3}$, ${\realone^{3N}}$, and ${\realone^{3 N N_c}}$, respectively, also span
the complex counterparts ${\complex^3}$, ${\complex^{3N}}$, and ${\complex^{3 N N_c}}$ of these spaces.

\subsection{Expressing structure in a basis of eigenvectors}
\label{section:eigenbasis_expansion}
We are finally ready to express the microstructure ${\kket{\psi}}$ of the crystal as a 
superposition of normal mode eigenvectors. 
I emphasize that doing so does not constitute an approximation, because the eigenvectors
are a complete basis set.
Therefore, let 
\begin{align}
\kket{\psi}
	& =\sum_{\{\bk,\bbk\}}\sum_{\mu} \Q_{\bk\mu} \, \kket{E_{\bk\mu}} 
 = \frac{1}{2}\sum_{\bk\mu} \Q_{\bk\mu} \, \kket{E_{\bk\mu}},
\label{eqn:expansion0}
\end{align}
where the coefficient ${\Q_{\bk\mu}=\Q_{\bbk\mu}}$ is  known as the {\em normal mode coordinate}
of mode ${\bk\mu}$; and, to reach the second expression, I used the fact that ${\kket{E_{\bbk\mu}}=\kket{E_{\bk\mu}}}$.
The normal mode coordinate is defined as
${\Q_{\bk\mu}\equiv\bbraket{E_{\bk\mu}}{\psi}}$. Therefore, using Eq.~\ref{eqn:eigenvector0} it 
can be expressed as 
\begin{align}
&\Q_{\bk\mu}
 = \frac{1}{\sqrt{2N_c}}\sum_{\bR} \left[\braket{\eig_{\bk\mu}}{\psi^\bR} e^{-i\bk\cdot\bR} + \braket{\eig_{\bbk\mu}}{\psi^\bR}e^{i\bk\cdot\bR}\right]
\label{eqn:q}
\end{align}
or as
\begin{align}
\Q_{\bk\mu}
& = \frac{1}{\sqrt{N_c}}\sum_{\bR} \abs{\braket{\eig_{\bk\mu}}{\psi^\bR}} \cos\left(\bk\cdot\bR+\vartheta_{\bk\mu}\right).
\end{align}
Equation~\ref{eqn:expansion0} is a simple mathematical statement. It states that the
structure of a crystal can be expressed as a sum of vectors that are elements 
of a complete basis of ${\realone^{3 N N_c}}$, and that this basis can be 
chosen to be a basis of standing lattice waves ${\kket{E_{\bk\mu}}}$ which, 
deep within the bulk, are almost exactly parallel to the crystal's normal mode eigenvectors.\\

\subsubsection{Time dependence of normal mode coordinates}
In the \templim limit, when the motion of the crystal's atoms
is a superposition of harmonic motions along the normal mode eigenvectors, ${\Q_{\bk\mu}(t)}$ 
has the simple form
${\Q_{\bk\mu}(t) =  A_{\bk\mu} 
\cos(\omega_{\bk\mu}t +\theta_{\bk\mu})}$, where $A_{\bk\mu}$ 
and ${\theta_{\bk\mu}}$ are determined by initial conditions.

In general, at finite temperature the only constraints on the set of observable frequencies of 
motions of the crystal along ${\kket{E_{\bk\mu}}}$ are the two that will be discussed in Sec.~\ref{section:discretizing}.
Before those constraints have been imposed, the most general expression for ${\Q_{\bk\mu}(t)}$ as a superposition
of motions with different frequencies is
\begin{align}
\Q_{\bk\mu}(t)  &= \sigma\int_{\realone}\dd{\omega} \tilde{\Q}_{\bk\mu}(\omega)e^{-i\omega t} 
\label{eqn:FTQ1}
\\
& = \sigma\int_{\realpos}\dd{\omega}\left[\tilde{\Q}_{\bk\mu}(\omega)e^{-i\omega t} + \tilde{\Q}_{\bk\mu}(-\omega)e^{i\omega t}\right]
\\
&=
2\sigma\int_{\realpos}\dd{\omega}\Re\left\{\tilde{\Q}_{\bk\mu}(\omega)e^{-i\omega t}\right\}
=\Q_{\bbk\mu}(t),
\nonumber
\end{align}
where I have used the fact that, because ${\Q_{\bk\mu}=\Q_{\bbk\mu}}$ is real, its FT with respect to time
satisfies
\begin{align*}
\tiQ_{\bk\mu}^*(\omega)=\tiQ_{\bk\mu}(-\omega)=\tiQ^*_{\bbk\mu}(\omega)=\tiQ_{\bbk\mu}(-\omega).
\end{align*}

We can express the time derivative of ${\Q_{\bk\mu}}$ as
\begin{align}
\dot{\Q}_{\bk\mu}(t) 
 & = -i\sigma\int_{\realone}\dd{\omega}\omega \tilde{\Q}_{\bk\mu}(\omega)e^{-i\omega t} 
\label{eqn:dQdtFT}
\\
 & = -i\sigma\int_{\realpos}\dd{\omega}\omega \left[\tilde{\Q}_{\bk\mu}(\omega)e^{-i\omega t} - \tilde{\Q}_{\bk\mu}^*(\omega)e^{i\omega t}\right]
\nonumber
\\
 & = 2\sigma\int_{\realpos}\dd{\omega} \omega\,\Im\left\{\tilde{\Q}_{\bk\mu}(\omega)e^{-i\omega t}\right\}.
\end{align}
Therefore its Fourier transform is 
\begin{align*}
\fourier_t\left[\dot{\Q}_{\bk\mu}\right](\omega)= -i\omega\tiQ_{\bk\mu}(\omega).
\end{align*}

Let ${\treversal}$ denote the time-reversal operator, which means that
we can write the time reversed normal mode coordinate of mode ${\bk\mu}$ as 
\begin{align}
\treversal\Q_{\bk\mu}(t)= \Q_{\bk\mu}(-t) 
& = \sigma\int_{\realone}\dd{\omega} \tilde{\Q}_{\bk\mu}(-\omega)e^{-i\omega t} 
\nonumber\\
\implies
\fourier_t\left[\treversal\Q_{\bk\mu}\right](\omega)&=\tilde{\Q}_{\bk\mu}(-\omega)
=\tiQ_{\bk\mu}^*(\omega).
\label{eqn:ft_treversal}
\end{align}
The FT of ${\treversal\dot{\Q}_{\bk\mu}}$ is 
\begin{align*}
\fourier_t\left[\treversal\dot{\Q}_{\bk\mu}\right](\omega)
&=
i\omega\tiQ_{\bk\mu}(-\omega)=i\omega\tiQ_{\bbk\mu}^*(\omega)
\\
&=\left(-i\omega\tiQ_{\bbk\mu}(\omega)\right)^* =\left(\fourier_t\left[\dot{\Q}_{\bk\mu}\right](\omega)\right)^*.
\end{align*}

To find an expression for ${\tiQ_{\bk\mu}(\omega)}$, let us substitute
Eq.~\ref{eqn:ft1} into  Eq.~\ref{eqn:q}, as follows:
\begin{widetext}
\begin{align}
\Q_{\bk\mu}(t) 
&=
\frac{\sigma}{\sqrt{2N_c}}
\sum_{\bk'}
\sum_\bR
\int_{\realone}\dd{\omega}
\left[
\braket{\eig_{\bk\mu}}{\tpsi_{\bk'}(\omega)}
e^{i(\bk'-\bk)\cdot\bR}
+
\braket{\eig_{\bbk\mu}}{\tpsi_{\bk'}(\omega)}
e^{i(\bk'+\bk)\cdot\bR}
\right]
e^{-i\omega t}
\nonumber
\\
&=
\sigma\sqrt{\frac{N_c}{2}}
\int_{\realone}\dd{\omega}
\left[
\braket{\eig_{\bk\mu}}{\tpsi_{\bk}(\omega)}
+
\braket{\eig_{\bk\mu}}{\tpsi_{\bbk}(\omega)}
+
\braket{\eig_{\bbk\mu}}{\tpsi_{\bbk}(\omega)}
+
\braket{\eig_{\bbk\mu}}{\tpsi_{\bk}(\omega)}
\right]
e^{-i\omega t},
\nonumber
\end{align}
\end{widetext}
where I have simplified using orthogonality relations
and ${\ket{\tpsi_\bk^*(\omega)}=\ket{\tpsi_{\bbk}(\bomega)}}$.
After replacing 
${\left(\ket{\eig_{\bk\mu}}+\ket{\eig_{\bbk\mu}}\right)/\sqrt{2}}$
with ${\ket{\neig_{\bk\mu}}}$ this becomes
\begin{align}
\Q_{\bk\mu}(t) 
=
\sigma\sqrt{N_c}
\int_{\realone}\dd{\omega}
\bigg[
&\braket{\neig_{\bk\mu}}{\tpsi_{\bk}(\omega)}
\nonumber \\
+
&\braket{\neig_{\bk\mu}}{\tpsi_{\bbk}(\omega)}
\bigg]
e^{-i\omega t}
\end{align}
Therefore, 
\begin{align}
\tiQ_{\bk\mu}(\omega)=\sqrt{N_c}
\bigg[
\braket{\neig_{\bk\mu}}{\tpsi_{\bk}(\omega)}
+
\braket{\neig_{\bk\mu}}{\tpsi_{\bbk}(\omega)}
\bigg]
\label{eqn:tiQdef}
\end{align}
and
\begin{align}
\fourier_t\left[\dot{Q}_{\bk\mu}\right](\omega)
=-i\omega\sqrt{N_c}\bigg[
&\braket{\neig_{\bk\mu}}{\tpsi_{\bk}(\omega)}
\nonumber\\
&+
\braket{\neig_{\bk\mu}}{\tpsi_{\bbk}(\omega)}
\bigg].
\label{eqn:tidQdef}
\end{align}

\subsection{Discretizing the time/frequency Fourier transform}
\label{section:discretizing}
Let $\tmax$ denote the time for which the crystal is observed or simulated, 
and I will refer to the interval during which it is observed as the
{\em observation interval}.
Let us assume that, for each ${\bk\mu}$, the function of time ${\Q_{\bk\mu}(t)}$ has been tapered
smoothly and rapidly to zero at the boundaries
of the observation interval. 
\subsubsection{Finite observation/simulation time}
\label{section:finite_time}
The finite observation time means that the smallest frequency that can be resolved is 
${\abs{\dd{\omega}}\equiv (\sigma^2 \tmax)^{-1}}$.
This finite resolution has consequences when evaluating the time average ${\expval{e^{i\omega t}}_t}$
of ${e^{i\omega t}}$, when it appears in an integral over $\omega$.
When it is not in an integrand, we can simply note that the time
average of a sinusoid vanishes unless its argument is zero, 
in which case its value and its time average are both one.
However, when the average appears inside an integral, 
the finite frequency resolution can be accounted for by
noting that,
in the small-${\omega}$ limit, ${\expval{e^{i\omega t}}_t}$ can be expressed as
\begin{align*}
\expval{e^{i\omega t}}_t &\equiv 
\frac{1}{\tmax}\int^{\tmax/2}_{-\tmax/2}(1+i\omega t)\dd{t}
\nonumber\\
&= \begin{cases}
1,&\; \omega\in
(-\abs{\dd{\omega}}/2,\abs{\dd{\omega}}/2),
\\
0, &\; \omega \notin (-\abs{\dd{\omega}}/2,\abs{\dd{\omega}}/2).
\end{cases}
\end{align*}
This implies that
\begin{align*}
\int_\realone\dd{\omega} \tilde{f}(\omega) \expval{e^{i\omega t}}_t
 =  \abs{\dd{\omega}} \tilde{f}(\omega).
\end{align*}

\subsubsection{Finite time resolution}
\label{section:finite_resolution}
Now suppose that the smallest time interval
that can be measured or simulated is ${\abs{\dd{t}}=\tmax/N_t}$, 
where ${N_t\in\integer}$. Then the largest frequency that
can be measured or simulated is ${\maxomega=(\sigma^2 \abs{\dd{t}})^{-1}= N_t/(\sigma^2 \tmax)=N_t\abs{\dd{\omega}}}$.
Therefore the 
numbers of frequencies and times that are sampled are equal.

Let us discretize the integral in the Fourier transform of ${f(t)}$ as
follows.
\begin{align}
\tilde{f}(\omega) 
& = \sigma\int_{\realone} f(t) e^{i\omega t} \dd{t} 
= \sigma \frac{\tmax}{N_t} \sum_t f(t) e^{i\omega t}
\nonumber \\
&= \frac{1}{N_t \sigma \abs{\dd{\omega}}} \sum_t f(t) e^{i\omega t}
\end{align}
To clear away unnecessary constants, 
let the discrete transform of ${f(t)}$ be defined as 
${\hat{f}(\omega)\equiv \sigma\abs{\dd\omega}\tilde{f}(\omega)}$, 
so that we have the discrete Fourier transform pair,
\begin{align*}
\hat{f}(\omega) & =\frac{1}{N_t} \sum_t f(t) e^{i\omega t} 
\;\;\text{and}\;\;
f(t)  = \sum_\omega \hat{f}(\omega) e^{-i\omega t},
\end{align*}
where the sums are over all sampled times and frequencies, respectively.
In particular, let ${\hat{\psi}_\bk(\omega)\equiv \sigma\abs{\dd\omega}\tpsi_{\bk}(\omega)}$
and let
${\hat{Q}_{\bk\mu}(\omega)\equiv \sigma\abs{\dd{\omega}}
\tilQ_{\bk\mu}(\omega)=\sqrt{N_c}\braket{\eig_{\bk\mu}}{\hpsi_\bk(\omega)}}$.
\\
\vspace{0.5cm}

\section{Distribution of kinetic energy in reciprocal spacetime}
\subsection{Kinetic energy expressed in of mode coordinates}
The kinetic energy of the crystal is 
\begin{align}
\K & = \frac{1}{2}\sum_{\bR j \alpha\beta} \dpsi^{\bR j \alpha *}g_{\alpha\beta} \dpsi^{\bR j \beta}
=
\frac{1}{2}\sum_{\bR j} \dbpsi^{\bR j *}\cdot\dbpsi^{\bR j}
\nonumber\\
&=
\frac{1}{2}
\sum_{\bR} \braket{\dpsi^\bR}{\dpsi^{\bR}}
= \frac{1}{2}\bbraket{\dpsi}{\dpsi} 
\label{eqn:q1}
\end{align}
By inserting Eq.~\ref{eqn:expansion0} into Eq.~\ref{eqn:q1}, this 
can be expressed as 
\begin{align}
\K(t)
= 
\frac{1}{2}
\sum_{\{\bk,\bbk\}}\sum_{\mu} 
\dot{\Q}_{\bk\mu}(t)^2,
\end{align}
which means that the time average of $\K$ divided by the total number of bulk cells
can be expressed as
\begin{align}
\frac{\expval{\K}_t}{N_c}=\frac{1}{2N_c}\sum_{\{\bk,\bbk\}}\sum_{\mu}
\expval{\dot{\Q}_{\bk\mu}^2}_t
= 
\frac{1}{4N_c}
\sum_{\bk\mu}
\expval{\dot{\Q}_{\bk\mu}^2}.
\end{align}
Then Eqs.~\ref{eqn:dQdtFT} and Eq.~\ref{eqn:tidQdef} 
can be used to express ${\expval{\dot{\Q}_{\bk\mu}^2}_t}$ as
\begin{widetext}
\begin{align*}
\expval{\dot{\Q}_{\bk\mu}^2}_t
&\equiv \frac{1}{\tmax}\int_\realone \dd{t} \dot{\Q}_{\bk\mu}(t)^2
=
-\frac{\sigma^2}{\tmax}\int_{\realone}\dd{\omega}\int_{\realone}\dd{\omega'}
\omega\omega'
\tiQ_{\bk\mu}(\omega)
\tiQ_{\bk\mu}(\omega')
\int_{\realone}e^{-i(\omega+\omega')t}\dd{t}
=\frac{1}{\tmax}\int_{\realone}\dd{\omega}
\omega^2
\tiQ_{\bk\mu}(\omega)
\tiQ_{\bk\mu}(-\omega)
\\
&=\frac{1}{\tmax}\int_{\realone}\dd{\omega}
\omega^2
\tiQ_{\bk\mu}(\omega)
\tiQ^*_{\bk\mu}(\omega)
=\frac{2 N_c}{\tmax}\int_\realone\dd{\omega}\omega^2
\left[
\braket{\tpsi_{\bk}(\omega)}{\neig_{\bk\mu}}
\braket{\neig_{\bk\mu}}{\tpsi_{\bk}(\omega)}
+
\braket{\tpsi_{\bk}(\omega)}{\neig_{\bk\mu}}
\braket{\neig_{\bk\mu}}{\tpsi_{\bbk}(\omega)}
\right].
\end{align*}
\end{widetext}
After replacing ${\omega}$ by ${-\omega}$ in the integral of the second term
in the parentheses, and using
${\tpsi^{j\alpha *}_\bk(\omega)=\tpsi^{j\alpha}_{\bbk}(\omega)=\tpsi^{j\alpha}_{\bbk}(-\omega)}$, 
this becomes
\begin{align*}
\expval{\dot{\Q}_{\bk\mu}^2}_t
=\frac{4 N_c}{\tmax}\int_\realone\dd{\omega}\omega^2
\braket{\tpsi_{\bk}(\omega)}{\neig_{\bk\mu}}
\braket{\neig_{\bk\mu}}{\tpsi_{\bk}(\omega)}
\end{align*}
Then, since ${\sum_\mu \dyad{\neig_{\bk\mu}}}$ is the identity in ${\complex^{3N}}$, 
and ${1/\tmax=\sigma^2\abs{\dd{\omega}}}$,
the time-averaged kinetic energy per primitive unit cell 
can be expressed as
\begin{align*}
\frac{\expval{\K}_t}{N_c} 
& = \sigma
\int_\realone\dd{\omega}
\tE^\K(\bk,\omega),
\end{align*}
where
\begin{align}
\tE^{\K}(\bk,\omega) 
&\equiv 
\sigma
\abs{\dd{\omega}}\omega^2
\braket{\tpsi_\bk(\omega)}
\nonumber
\\
&= \frac{\omega^2}{\sigma\abs{\dd{\omega}}}\braket{\hpsi^{j\alpha}_{\bk}(\omega)}.
\label{eqn:ekin}
\end{align}
Since ${\tE^{\K}(\bk,\omega)}$ vanishes at values of $\omega$ that
are not integer multiples of ${\abs{\dd{\omega}}\equiv 2\pi/\tmax}$, 
we can replace the integral by a sum and express Eq.~\ref{eqn:ekin}
as
\begin{align}
\frac{\expval{\K(t)}_t}{N_c}  
= 
 \sum_{\bk}\sum_{\omega} \hE^{\K}(\bk,\omega).
\end{align}
where
\begin{align}
\hE^{\K}(\bk,\omega) \equiv 
\omega^2
\braket{\hpsi_\bk(\omega)}.
\label{eqn:ekin2}
\end{align}
Therefore ${\hE^{\K}(\bk,\omega)}$ is
the contribution of point ${(\bk,\omega)}$ in reciprocal spacetime
to the average kinetic energy per unit cell.

\subsection{Expressing structure in a basis of complex waves}
\label{section:complex_waves}
In the previous section we expressed structure in terms of a set
of normal mode coordinates, ${\Q_{\bk\mu}=\Q_{\bbk\mu}}$. However, 
at finite temperature each vibrational excitation does not involve the motion
of every atom in the crystal. Therefore it has a finite size and is a travelling
wave or, more accurately, a travelling wave packet. 
This makes it important to distinguish between the two counterpropagating 
waves, $\bk\mu$ and $\bbk\mu$, that contribute to each standing wave ${\Q_{\bk\mu}\kket{E_{\bk\mu}}}$.

In this section I introduce the {\em complex mode coordinates} ${\{Q_{\bk\mu}\}}$,
for which ${Q_{\bk\mu}\neq Q_{\bbk\mu}}$, in general.
As their name suggests, and unlike the {\em normal mode coordinates} ${\Q_{\bk\mu}}$, they are not real, in general.
Therefore it is in this section that we truly begin to use complex vector spaces. 

Let us begin with the definition
\begin{align}
\Eig_{\bk\mu}^{\bR j \alpha}\equiv \frac{1}{\sqrt{N_c}}\eig_{\bk\mu}^{j \alpha}e^{-i\bk\cdot\bR}
=
\Eig_{\bbk\mu}^{\bR j \alpha *} \in\complex,
\label{eqn:complex_eigenvector}
\end{align}
which implies that
	${E_{\bk\mu}^{\bR j \alpha}=\left[\Eig_{\bk\mu}^{\bR j \alpha}+\Eig_{\bbk\mu}^{\bR j \alpha}\right]/\sqrt{2}}$.
Now we can define the {\em complex mode eigenvector}, or simply {\em mode eigenvector}, of mode ${\bk\mu}$ to be
\begin{align}
\kket{\Eig_{\bk\mu}}\equiv\sum_{\bR j \alpha}\Eig_{\bk\mu}^{\bR j \alpha}\kket{\bR j \alpha}\in\complex^{3 N  N_c},
\label{eqn:eigenwave}
\end{align}
which implies the following relations:
\begin{align*}
	\kket{E_{\bk\mu}}&=\frac{1}{\sqrt{2}}\left[\kket{\Eig_{\bk\mu}}+\kket{\Eig_{\bbk\mu}}\right] 
\\
	\Q_{\bk\mu}&=
	\frac{1}{\sqrt{2}}\left[\bbraket{\Eig_{\bk\mu}}{\psi}+\bbraket{\Eig_{\bbk\mu}}{\psi}\right]
	=\sqrt{2}\,\eta\left(\kket{\Eig_{\bk\mu}},\kket{\psi}\right)
\end{align*}
Note that ${\Q_{\bk\mu}}$ is a factor of ${\sqrt{2}}$ larger than 
${\eta(\left(\kket{\Eig_{\bk\mu}},\kket{\psi}\right)}$ because there are twice as many
wavevectors contributing to the basis of complex mode eigenvectors
as there are wavevector pairs contributing to the basis of normal mode eigenvectors.

Equation~\ref{eqn:expansion0} can now be expressed as
\begin{align}
\kket{\psi} 
&= 
\frac{1}{4}\sum_{\bk\mu}
\left[\bbraket{\Eig_{\bk\mu}}{\psi} + \bbraket{\Eig_{\bbk\mu}}{\psi}\right]
\left[\kket{\Eig_{\bk\mu}} + \kket{\Eig_{\bbk\mu}}\right]
\nonumber
\\
&= 
	\frac{1}{2}\sum_{\bk\mu}
\left[\bbraket{\Eig_{\bk\mu}}{\psi} + \bbraket{\Eig_{\bbk\mu}}{\psi}\right]
\kket{\Eig_{\bk\mu}}
\nonumber
\\
&= 
\sum_{\bk\mu}
\eta\left(\kket{\Eig_{\bk\mu}},\kket{\psi}\right)
\kket{\Eig_{\bk\mu}}
= 
\sum_{\bk\mu}
	Q_{\bk\mu}
\kket{\Eig_{\bk\mu}},
\label{eqn:mode_coord}
\end{align}
where ${Q_{\bk\mu}\equiv\bbraket{\Eig_{\bk\mu}}{\psi}}$, 
which implies that ${Q_{\bk\mu}^*=Q_{\bbk\mu}}$,
and that 
\begin{align*}
\Re\left\{Q_{\bk\mu}\right\}=\eta\left(\kket{\Eig_{\bk\mu}},\kket{\psi}\right) = \frac{\Q_{\bk\mu}}{\sqrt{2}}.
\end{align*}
Let us use superscripts ${\Re}$ and ${\Im}$ to denote
real and imaginary parts, respectively, so that
\begin{align*}
Q_{\bk\mu}(t) &= Q^{\Re}_{\bk\mu}(t)+iQ^{\Im}_{\bk\mu}(t)
=\frac{\Q_{\bk\mu}(t)}{\sqrt{2}}+iQ^{\Im}_{\bk\mu}(t), 
\end{align*}
The only constraint imposed, either implicitly or explicitly, on the imaginary part 
of ${Q_{\bk\mu}}$ is ${Q^*_{\bk\mu}=Q_{\bbk\mu}}$. Therefore, 
it can be chosen to choose a useful purpose and we will return
to discussing useful choices of it in Sec.*.

\subsubsection{Harmonic approximation}
\label{section:harmonic}
When the only active mode is ${\bk\mu}$, each atom undergoes
simple harmonic motion with the same frequency, ${\omega_{\bk\mu}}$.
Therefore, when the amplitude of mode $\bk\mu$ is sufficiently
small, the potential
energy is well approximated by the harmonic expression,  
\begin{align}
\E_2 &\equiv \frac{1}{2}\sum_{\bR j} m_j\omega_{\bk\mu}^2\abs{\bu_\bR}^2 
=\frac{1}{2}\sum_\bR \omega_{\bk\mu}^2\abs{\bpsi^\bR}^2 
\nonumber\\
&= \frac{1}{2}\omega_{\bk\mu}^2 Q_{\bk\mu}^* Q_{\bk\mu}
= \frac{1}{2}\omega_{\bk\mu}^2 Q_{\bbk\mu} Q_{\bk\mu},
\end{align}
where a route similar to the one taken in Eq.~\ref{eqn:q1} was used to reach the final expression 
in terms of the mode coordinates.
When all modes are active, the harmonic
approximation to the energy is
\begin{align}
\E \approx \K+ \E_2 
&=
\frac{1}{2}\sum_{\bk\mu}\left[\dot{Q}_{\bbk\mu}\dot{Q}_{\bk\mu} + \omega^2_{\bk\mu} Q_{\bbk\mu}Q_{\bk\mu}\right]
\nonumber \\
&=
\frac{1}{2}\sum_{\bk\mu}\left[P_{\bk\mu}P_{\bbk\mu} + \omega^2_{\bk\mu} Q_{\bbk\mu}Q_{\bk\mu}\right],
\label{eqn:E}
\end{align}
where ${P_{\bq\mu}\equiv \dot{Q}_{\bq\mu}^*= \dot{Q}_{\bbq\mu}}$ is the momentum conjugate
to ${Q_{\bq\mu}}$.

Note that
Eq.~\ref{eqn:E} simplifies to ${\K+\E_2 = \sum_{\bk\mu}\omega_{\bk\mu}^2 Q_{\bbk\mu} Q_{\bk\mu}}$
when the system is harmonic, 
because ${\dot{Q}_\bk^\mu=-i\omega_{\bk\mu}Q_{\bk}^\mu}$. However, referring
to Eq.~\ref{eqn:q}, we see that, in general (i.e., at finite $T$), 
the oscillation along eigenvector ${\kket{E_{\bk\mu}}}$ has contributions from a continuous
range of frequencies. In the \templim limit, Eq.~\ref{eqn:q} must reduce to the form
${Q_{\bk\mu}\sim \cos \omega_{\bk\mu}t}$. Therefore, the Fourier transform ${\tilQ_{\bk\mu}(\omega)}$ of
${Q_{\bk\mu}(t)}$  must become more and more sharply peaked at ${\omega=\omega_{\bk\mu}}$ as the ${T\to 0}$ limit is approached.

\subsubsection{Perturbation theory of interacting phonons}
The expression for $\K$ does not change at finite $T$, but the expression for the potential
energy does. However, the set of mode coordinates ${\{Q_{\bq\mu}\}}$ provides a complete
specification of the positions of all atoms in the crystal via Eq.~\ref{eqn:mode_coord}.
Therefore  the potential energy can be expressed as a function of all mode coordinates
and then Taylor expanded about mechanical equilibrium (${Q_{\bq\mu} = 0, \;\forall\, \bk\mu}$).
The harmonic term in this expansion must equal ${\E_2}$, because all other terms
vanish as the $Q$'s become vanishingly small, which is the limit
in which Eq.~\ref{eqn:E} becomes exact.

When displacements from equilibrium
are large enough that anharmonic contributions to the energy
are relevant, the energy may be expressed as 
\begin{align}
&\E  = \K+
\E_2 + 
\sum_{\substack{\mu_1\mu_2\mu_3\\ \bq_1\bq_2\bq_3}} 
\tensor*{\Phi}{*_{\mu_1}^{\bq_1}_{\mu_2}^{\bq_2}_{\mu_3}^{\bq_3}}
Q_{\bq_1\mu_1} Q_{\bq_2\mu_2} Q_{\bq_3\mu_3}
\nonumber \\
&+ \sum_{\substack{\mu_1\mu_2\mu_3\mu_4\\ \bq_1\bq_2\bq_3\bq_4}}  
\tensor*{\Phi}{*_{\mu_1}^{\bq_1}_{\mu_2}^{\bq_2}_{\mu_3}^{\bq_3}_{\mu_4}^{\bq_4}}
Q_{\bq_1\mu_1} Q_{\bq_2\mu_2} Q_{\bq_3\mu_3} Q_{\bq_4\mu_4}
+\order{Q^5},
\label{eqn:energy_expansion}
\end{align}
where ${Q^p}$ represents the product
of any $p$ mode coordinates; and
each coefficient $\Phi$ with $l$ subscripts and $l$ superscripts is ${1/l!}$ times 
the $l^\text{th}$ partial derivative of the potential energy with
respect to the mode coordinates identified by those indices, 
and evaluated at equilibrium, where all $Q$'s vanish. 
The anharmonic part of the potential energy, which I will denote by ${\DE}$, 
contains an infinite number
of terms, but is usually truncated after the ${Q^3}$ or ${Q^4}$ terms
in practical usage. 

I will denote the sets of all mode coordinates and momenta by
${\bQ\equiv(Q_{\bk_1,\mu_1},Q_{\bk_2,\mu_2},\cdots)}$
and ${\bP\equiv(P_{\bk_1,\mu_1},P_{\bk_2,\mu_2},\cdots)}$, respectively.
This allows Eq.~\ref{eqn:energy_expansion} to be expressed as
\begin{align}
\E(\bQ,\bP) & = \K(\bP) + \E_2(\bQ)+\DE(\bQ).
\end{align}
It is important that $\DE$ does not
depend on ${\bP}$. If it did, ${\dot{Q}_{\bbk\mu}}$ would cease
to be the momentum conjugate to ${Q_{\bk\mu}}$ at finite $T$ whenever
${\DE}$ is not negligible.

\subsubsection{Kinetic energy density in reciprocal spacetime from correlation functions}
\label{section:correlation}
The mass-weighted velocity-velocity correlation
function (mVVCF) of the $j^\text{th}$ atom in the primitive unit cell
is defined as
\begin{align*}
C^j(\bR,t) \equiv m_j \expval{ \dbu^{\bR_0\,j}(t_0)\bcdot\dbu^{\bR_0+\bR\, j}(t_0+t)}_{\bR_0, t_0},
\end{align*}
where ${\expval{\;}_{\bR_0, t_0}}$ denotes an average over all $\bR_0\in\bravaisbulk$ and over 
all initial times $t_0$.
I will now derive a simple relationship between the discrete distribution of vibrational energy
in reciprocal spacetime, $\hE(\bk,\omega)$, and the discrete Fourier
transform, ${\hC(\bk,\omega)}$, of
${C(\bR,t)\equiv \sum_j C^j(\bR,t)}$, which can 
also be expressed as
\begin{align*}
C&(\bR,t) 
 = \expval{\braket{\dpsi_{\bR_0}(t_0)}{\dpsi_{\bR_0+\bR}(t_0+t)}}_{\bR_0, t_0}.
\end{align*}
I begin by replacing ${\bra{\dpsi_{\bR_0}(t_0)}}$ and ${\ket{\dpsi_{\bR_0+\bR}(t_0+t)}}$
with their discrete Fourier transforms with respect to both space and time, 
and I also make the averages over $\bR_0$ and $t_0$ more explicit, as follows:
\begin{widetext}
\begin{align*}
C(\bR,t) 
& = \frac{1}{N_c}\sum_{\bR_0} \frac{1}{N_t}\sum_{t_0}\sum_{\bk\omega}\sum_{\bk'\omega'}
\omega\omega'\braket{\hpsi_\bk(\omega)}{\hpsi_{\bk'}(\omega')}
e^{i(\bk'-\bk)\cdot\bR_0}e^{-i(\omega'-\omega)t_0}e^{i(\bk'\cdot\bR-\omega't)}
\nonumber\\
& = \sum_{\bk\omega}\sum_{\bk'\omega'} \omega\omega' \braket{\hpsi_\bk(\omega)}{\hpsi_{\bk'}(\omega')}
\left(\frac{1}{N_c}\sum_{\bR_0} e^{i(\bk'-\bk)\cdot\bR_0}\right)
\left(\frac{1}{N_t}\sum_{t_0} e^{-i(\omega'-\omega)t_0}\right)
e^{i(\bk'\cdot\bR-\omega't)}
\end{align*}
\end{widetext}
Note that the quantities in parentheses  can be replaced with
${\delta_{\bk\bk'}}$ and ${\delta_{\omega\omega'}}$; therefore,
\begin{align*}
C(\bR,t) 
& = \sum_{\bk\omega} \omega^2\braket{\hpsi_\bk(\omega)}{\hpsi_{\bk}(\omega)} e^{i(\bk\cdot\bR-\omega t)}
\nonumber\\
& = \sum_{\bk\omega} \hC(\bk,\omega) e^{i(\bk\cdot\bR-\omega t)}
\end{align*}
where 
${\hC(\bk,\omega)=\omega^2 \braket{\hpsi_\bk(\omega)}{\hpsi_{\bk}(\omega)}}$
is the discrete Fourier transform of ${C(\bR,t)}$.
It follows from Eq.~\ref{eqn:ekin2} that
\begin{align}
\hE^{\K}(\bk,\omega) 
&\equiv \frac{1}{2}
\left[
\hC(\bk,\omega)+\hC(\bbk,\omega)
\right].
\end{align}
and, since ${\braket{\hpsi_\bk(\omega)}=\braket{\hpsi^*_{\bk}(\omega)}}$
implies that 
${\hC(\bbk,\bar{\omega})=\hC(\bk,\omega)}$ 
and
${\hC(\bbk,\omega)=\hC(\bk,\bar{\omega})}$, 
the average kinetic energy per unit cell can be expressed as
\begin{align}
\frac{\expval{\K(t)}}{N_c} = \sum_\bk\sum_\omega \hC(\bk,\omega),
\label{eqn:correlation_energy}
\end{align}
where the sum over $\omega$ is not restricted to positive values.\\

\subsubsection{Mode-projected correlation functions}
\label{section:mode_projection}
By expressing the identity of ${\complex^{3N}}$
as ${\sum_{\bk\mu}\dyad{\eig_{\bk\mu}}}$ 
or as ${\sum_{\bR\mu}\dyad{\wannier_{\bR\mu}}}$,
we can write ${C(\bR,t)}$ in the following forms.
\begin{align}
& C(\bR,t) 
\nonumber\\
& = 
\sum_{\bk\mu}
  \expval{\braket{\dpsi_{\bR_0}(t_0)}{\eig_{\bk\mu}}\braket{\eig_{\bk\mu}}{\dpsi_{\bR_0+\bR}(t_0+t)}}_{\bR_0, t_0}
\nonumber\\
& = 
\sum_{\bR'\mu}
  \expval{\braket{\dpsi_{\bR_0}(t_0)}{\wannier_{\bR'\mu}}\braket{\wannier_{\bR'\mu}}{\dpsi_{\bR_0+\bR}(t_0+t)}}_{\bR_0, t_0}
\end{align}
This means that, for example, we can calculate the contribution to the reciprocal spacetime
distribution of the 
kinetic energy of motion parallel to each $\ket{\eig_{\bk\mu}}$ separately: It is the Fourier transform,
${\hC_{\bk\mu}(\bk,\omega)}$, 
of the {\em mode projected correlation function}, 
\begin{align}
C_{\bk\mu}(\bR,t)
\equiv 
\bigg\langle&\braket{\dpsi_{\bR_0}(t_0)}{\eig_{\bk\mu}}
\nonumber \\
\times &\braket{\eig_{\bk\mu}}{\dpsi_{\bR_0+\bR}(t_0+t)}
\bigg\rangle_{\bR_0, t_0}.
\label{eqn:mode_correlation}
\end{align}
${\hC_{\bk\mu}(\bk',\omega)}$ 
is non-zero at all wavevectors ${\bk'\neq\bk}$, in general,
because cell eigenvectors at different wavevectors
are not orthogonal to one another. Therefore, each
cell eigenvector ${\ket{\eig_{\bk\mu}}}$ 
is a superposition of the cell eigenvectors, ${\left\{\ket{\eig_{\bk'\mu}}\right\}_{\mu=1}^{3N}}$,
at any wavevector ${\bk'}$. This means that, as a function of
${\omega}$ at a fixed wavevector ${\bk'}$, 
${\hC_{\bk\mu}(\bk',\omega)}$ has contributions from 
all modes ${\{\bk'\mu:1\leq \mu\leq 3N\}}$.
Therefore, we will usually focus on the 
value of ${\hC_{\bk\mu}(\omega)\equiv \hC_{\bk\mu}(\bk',\omega)\eval_{\bk'=\bk}}$, 
after suitably normalizing it.
In the low $T$ limit, each mode-projected
spectrum  ${\hC_{\bk\mu}(\omega)}$ is a single sharp peak
at ${(\bk,\omega_{\bk\mu})}$. As $T$
increases these peaks are expected to broaden, to change shape, 
and to shift in frequency.


\section{Phonon sizes: From standing waves to wave packets and 
the quasiparticle gas}
[This section is far from complete and under heavy construction.]

The focus of the present work is on the classical wave theory of phonons.
Its purposes are to lay out this theory's mathematical
foundations in a more general and comprehensive form than can 
be found in most existing literature, and to derive a classically-exact
expression for the decomposition of a crystal's kinetic energy into
contributions ${\hE^\K(\bk,\omega)}$
from a set of sampled points ${(\bk,\omega)}$ in reciprocal spacetime.

At each point \wk~the energy 
can be decomposed further by expressing it as the sum,
\begin{align*}
\hE^\K(\bk,\omega)= \sum_{\mu=1}^{3N}\hE^\K_{\bk\mu}(\omega),
\end{align*}
of as many contributions ${\hE^\K_{\bk\mu}(\omega)}$ as there
are degrees of freedom in each primitive unit cell of the crystal.
In Sec.~\ref{section:mode_projection} I chose to decompose it
into contributions from motions along the \templim~normal mode cell eigenvectors, ${\ket{\eig_{\bk\mu}}}$, 
but I could have chosen to decompose it into contributions from motions along any set of ${3N}$ vectors
that spans ${\realone^{3N}}$.

For example, at each temperature and each wavevector $\bk$, there must exist an orthonormal
basis, ${\basis\equiv\left\{\ket{e_{\bk\mu}}\right\}_{\mu=1}^{3N}}$,
of ${\realone^{3N}}$ that minimizes the sum of the second central moments of the ${3N}$ contributions
to ${\hE^\K(\bk,\omega)}$, i.e., that minimizes
\begin{align*}
\cost\left[\basis\right]\equiv\sum_{\mu=1}^{3N} \frac{1}{\N_{\bk\mu}}\int_{\realpos}\left(\omega - \bar{\omega}_{\bk\mu}\right)^2 \hE^\K_{\bk\mu}(\omega)\dd{\omega}
\end{align*}
with respect to $\basis$, 
where ${\N_{\bk\mu}}$ is some chosen normalization constant, such as unity, or 
\begin{align*}
\N_{\bk\mu}\equiv
\big\|\hE_{\bk\mu}\big\|_1\equiv\int_{\realpos}\hE^{\K}_{\bk\mu}(\omega)\dd{\omega},
\end{align*}
or any other judicious and appropriate choice, and
\begin{align*}
\bar{\omega}_{\bk\mu}\equiv \big\|\hE_{\bk\mu}\big\|_1^{-1}\int_{\realpos} \omega \,\hE^{\K}(\bk,\omega)\dd{\omega}.
\end{align*}
Choosing basis $\basis$ such that the functions ${\hE^\K_{\bk\mu}(\omega)}$
are localized would reduce the degree of overlap between them.
Therefore it is likely to reduce the rate at which the motions along these vectors
\emph{resonantly} exchange energy with one another. 
However, resonance is not necessary
for interactions to be strong, as discussed and demonstrated
in Ref.~\onlinecite{coiana_2023}. Therefore the basis $\basis$
that minimizes $\cost$ may not define a set of mutually-orthogonal
motions that exchange energy with one another more slowly.
Motions along different elements of ${\basis}$ may interact
with one another more strongly, on average,
than motions along vectors for which the set ${\{\hE_{\bk\mu}\}_{\mu=1}^{3N}}$
is less localized.



One can also decompose ${\hE^\K(\bk,\omega)}$ into contributions from motions
along the eigenvectors of a self-consistent temperature-renormalized
dynamical matrix.  
In other words, along a set of mutually-orthogonal vectors that have been
chosen such that the time-averaged mean-field interaction between small-amplitude motion
along each one, and the rest of the crystal's vibrations, vanishes
at first order in its amplitude.



\subsection{Wave perspective versus scattering perspective} 
\label{section:perspective}
Whether it is most appropriate to treat phonons as waves or as quasiparticles 
depends on their sizes. By the `size' of a phonon I mean its linear dimensions
in all directions.
Along its axis of propagation its size is quantified by its coherence length, which I 
will discuss briefly in Sec.~\ref{section:coherence}.
In the planes perpendicular to its wavevector, its size is quantified by the linear dimensions
of the part of the crystal that it perturbs as it passes by, which 
can be quantified by its {\em coherence radius}.

At finite $T$, when the sizes of phonons
are orders of magnitude smaller than the size of the crystal, 
each phonon can be regarded as a quasiparticle that is born in a scattering event, 
travels until it collides with, and scatters from, other phonon
quasiparticles, and eventually dies.
Put another way, when it is meaningful to regard each phonon as having a well defined position at
any given instant within its lifetime, the scattering perspective is useful. When it is not, the wave perspective
is usually more appropriate.

\subsubsection{Correlation and coherence lengths and times}
\label{section:coherence}
Given a correlation function, such as ${C(\bR,t)}$ or
${C_{\bk\mu}(\bR,t)}$, I use the term
{\em correlation length} to mean its characteristic decay
length, either at ${t=0}$ or when it is integrated over $t$.
We will not need to be specific about how a characteristic
decay length is defined; it could be the (direction-dependent)
magnitude of ${\bR}$ at
which the value of ${\abs{C_{\bk\mu}(\bR,t)/C_{\bk\mu}(0,t)}}$
becomes lower than a specified value or, for an
exponential decay, ${C_{\bk\mu}(\bR,t)\sim e^{-\gamma \bR}}$, 
it could be $1/\gamma$.



A {\em correlation time} is the characteristic decay time
of a correlation function at $\bR=0$, or when 
it is summed over all ${\bR\in\bravaisbulk}$, and a {\em coherence time} is 
the correlation time when ${\bR\equiv \left(2\pi/\abs{\bk}\right)\hat{\bk}}$, where
${\hat{\bk}\equiv\bk/\abs{\bk}}$.

\subsection{Inadequacy of mode coordinates within scattering picture of phonon-phonon interactions}
When phonons are small enough that it is appropriate to describe their interactions as discrete
spatially-localized scattering events, neither the mode coordinate ${Q_{\bk\mu}}$ 
nor the contribution 
${\frac{1}{2}\left[\dot{Q}_{\bbk\mu}\dot{Q}_{\bk\mu} + \omega^2_{\bk\mu} Q_{\bbk\mu}Q_{\bk\mu}\right]}$,
of mode ${\bk\mu}$ to the total energy, is a satisfactory
measure of the mode ${\bk\mu}$'s prevalence within the crystal.
This is because ${\abs{Q_{\bk\mu}}}$ is the {\em net}
amplitude of ${\bk\mu}$ phonons in the crystal as a whole. If there are many independent ${\bk\mu}$
phonons in different parts of the crystal, their phases will all be different and
almost all of their contributions to ${Q_{\bk\mu}}$ will cancel one another out.
Therefore, it might be useful to build a scattering theory 
in which the positions and sizes/inertias of phonons are recognized.

\subsubsection{Basis of Wannier vectors}
${Q_{\bk\mu}}$ is simply the overlap between
the crystal's displacement from equilibrium
and the eigenvector of mode $\bk\mu$. 
By a slight manipulaton of Eq.~\ref{eqn:q}, we can
also express it as ${\sqrt{N_c}}$ times
the projection onto cell eigenvector ${\ket{\eig_{\bk\mu}}}$
of the Fourier transform of ${\ket{\psi_{\bR}(t)}}$ with
respect to $\bR$, i.e.,
\begin{align}
Q_{\bk\mu}(t) = \sqrt{N_c}\bra{\eig_{\bk\mu}}\left(\frac{1}{N_c}
\sum_\bR \ket{\psi_\bR(t)} e^{-i\bk\cdot\bR}\right).
\label{eqn:wannier1}
\end{align}
By assumption, the cell eigenvectors do not vary in the bulk; however, 
because
they vary quasicontinuously with wavevector $\bk$, we can
Fourier transform them with respect to $\bk$ to find 
a set of vectors ${\{\ket{\wannier_{\bR\mu}}:\bR\in\bravaisbulk\}}$
that are localized in real space. I refer to these as {\em Wannier vectors}
because they are the phonon counterparts of the {\em Wannier functions} that
appear in the band theory of electrons. 
Each one is defined with reference to a particular
cell $\Omega_\bR$ and a particular branch $\mu$
as
\begin{align*}
\ket{\wannier_{\bR\mu}} & \equiv \sum_\bk\ket{\eig_{\bk\mu}}e^{i\bk\cdot\bR} 
\iff \ket{\eig_{\bk\mu}}  = \frac{1}{N_c} \sum_\bR \ket{\wannier_{\bR\mu}}e^{-i\bk\cdot\bR}
\end{align*}
Substituting the expression for ${\ket{\eig_{\bk\mu}}}$ into Eq.~\ref{eqn:wannier1} gives
\begin{align}
Q_{\bk\mu}(t) & = \frac{1}{N_c}\sum_\bR 
\bigg(
\frac{1}{\sqrt{N_c}} \sum_{\bR'} \bra{\wannier_{\bR'\mu}} e^{i\bk\cdot\bR'}
\bigg)
\ket{\psi_\bR} e^{-i\bk\cdot\bR}
\nonumber \\
& = 
\expval{
\frac{1}{\sqrt{N_c}}\sum_{\dbR} \braket{\wannier_{\bR+\dbR\,\mu}}{\psi_\bR} e^{i\bk\cdot\dbR}
}_\bR,
\label{eqn:wannier2}
\end{align}
where I have used the substitution ${\bR'=\bR+\dbR}$.

Now let us define {\em cell-normalized mode coordinate}
${\barq_{\bk\mu}\equiv Q_{\bk\mu}/\sqrt{N_c}}$, which allows
us to express the energy per unit cell of mode ${\bk\mu}$ as
${\frac{1}{2}\dot{\barq}^*_{\bk\mu}\dot{\barq}_{\bk\mu}
=\frac{1}{2} \dot{Q}^*_{\bk\mu}\dot{Q}_{\bk\mu}/N_c}$.
Then Eq.~\ref{eqn:wannier2} can be used to
express its Fourier transform 
with respect to ${\bk}$ as
\begin{align*}
\barq_{\bR\mu}
\equiv
\expval{\braket{\wannier_{\bR_0+\bR\,\mu}}{\psi_{\bR_0}}}_{\bR_0}=
\expval{\braket{\wannier_{\bR_0\mu}}{\psi_{\bR_0+\bR}}}_{\bR_0}.
\end{align*}
This is simply the average overlap of the structures of bulk cells
with Wannier vectors referenced to cells displaced from them by $\bR$, or
the average overlap of Wannier vectors with the structures
of cells displaced from them by ${\bR}$.
It is straightforward to show that the kinetic energy per unit cell
can be expressed in either of the following forms.
\begin{align}
\frac{\K(t)}{N_c} 
= \frac{1}{2}\sum_{\bk\mu} \dot{\barq}^*_{\bk\mu}\dot{\barq}_{\bk\mu}
= \frac{1}{2}\sum_{\bR\mu} \dot{\barq}^*_{\bR\mu}\dot{\barq}^*_{\bR\mu}.
\end{align}


\subsubsection{Suggestions for further development of phonon theory}
One could define the {\em local mode coordinate},
${q_{\bk\mu}(\bR,t)\equiv \braket{\eig_{\bk\mu}}{\psi_\bR(t)}e^{-i\bk\cdot\bR}}$,
whose average over all ${\bR}$ is ${\barq_{\bk\mu}(t)}$.
From there one could try to identify clusters of cells,
$\Omega_\bR$, in which the modulus of
${q_{\bk\mu}(\bR,t)}$ is large and all cells in each cluster are participants in the
same phonon's motion. Then one could attempt to characterize each cluster, which is 
a wave packet of finite size, as a point particle with suitably-chosen definitions of
quantities such as its position, inertia, and linear momentum. Its linear momentum is likely
to be proportional to the group velocity of the wave packet, and is likely to differ from the
sum, over all cells in the cluster, of their contributions, ${\dot{q}_{\bbk\mu}(\bR,t)}$, to the mode momentum.

After deducing an appropriate way to characterize each phonon particle, 
a statistical theory of scattering could be built, which 
acknowledges the localized
nature of phonons at finite $T$, the inhomogeneity of their spatial distribution, 
and the nonuniformity of their sizes and inertias.

However, I am restricting the scope of the present work to describing phonons
when it is most appropriate to treat them as waves, rather than particles.
This is often the case in the \templim~limit, or when studying lattice waves in nanoscale or nanostructured materials,  
or when analysing the results of atomistic simulations. 
It is the case when one is studying phonon dynamics on length scales small 
enough that the presence of a $\bk\mu$ phonon in the region of the crystal being studied implies
that all of the primitive cells in that region are participating in the phonon's vibration
and propagation.  When this is the case, it follows that no
more than one ${\bk\mu}$ phonon exists within the region of interest because
the superposition of two $\bk\mu$ phonons would be equivalent to a single $\bk\mu$ phonon.

\section{Summary}
This is a work-in-progress whose completion will be gradual. Some sections are only partly written, and
there are likely to be mistakes throughout the manuscript.
I am placing it into 
the public domain at this time in the hope that others will find mistakes in it and
notify me of them, and/or complete the theory themselves.

The purpose of this work is to present the theory of vibrations in crystals
in a more general form than is commonly found in textbooks and journals. It is hoped that, 
when it eventually reaches completion, the many mistakes within it will have been fixed, 
and it can serve as a self-contained general reference for those simulating vibrations in crystals
and/or developing the theory of vibrations in crystals further.

\appendix
\appendixpage
\section{Fourier transforms}
\subsection{Continuous Fourier transforms}
\label{section:fourier}
My starting point is the unitary form of the Fourier transform (FT), with ${\sigma\equiv 1/\sqrt{2\pi}}$,
and with different signs in the imaginary exponents for the transforms with respect to space and time, i.e., 
the FT of an arbitrary function ${f(\bR,t)}$ with respect to $t$ is
\begin{align*}
\fourier_t[f](\bR,\omega)\equiv \sigma\int_{\realone}\dd{t} f(\br,t)e^{i\omega t},
\end{align*}
its FT with respect to $\br$ is 
\begin{align*}
\fourier_s[f](\bk,t)\equiv \sigma^3\int_{\realone^3} \dd[3]{r} f(\br,t) e^{-i\bk\cdot\br},
\end{align*}
and its FT with respect to both ${\br}$ and ${t}$ is 
\begin{align*}
\tilde{f}(\bk,\omega) \equiv
\fourier[f](\bk,\omega)\equiv
\fourier_s\left[\fourier_t [f]\right](\bk,\omega)
=
\fourier_t\left[\fourier_s [f]\right](\bk,\omega).
\end{align*}
I adopt different conventions for the signs of the imaginary exponents in the integrands
of ${\fourier_t[f]}$ and ${\fourier_s[f]}$
so that the FT with respect to both $\br$ and $t$, ${\tilde{f}(\bk,\omega)}$, is the coefficient
of a complex wave ${e^{i(\bk\cdot\br-\omega t)}}$ that travels in the direction of ${\bk}$ when
${\omega}$ is positive.

\subsubsection{Orthogonality relations for continuous waves}
Complex waves are mutually orthogonal functions of ${(\br,t)}$, which means that 
they satisfy
\begin{align*}
\int_{\realone}
\dd{\omega}\int_{\realone^3} \dd[3]{k}
\left(e^{i(\bk\cdot\br-\omega t)}\right)^*
&e^{i(\bk\cdot\br'-\omega t')} 
\\
&= \left(2\pi\right)^4\delta^{(3)}(\br-\br')\delta(t-t'),
\end{align*}
The geometric reason for taking the complex conjugate of one
of the waves in the orthogonality relation
will be  discussed in Sec.~\ref{section:vectors}.

Individually, the temporal and spatial parts of the complex waves
satisfy the following orthogonality relations, which are proved
in many undergraduate mathematics textbooks~\cite{riley_hobson_bence, boas}.
\begin{align}
\int_\realone\dd{\omega}e^{\pm i\omega(t-t')} &= 2\pi\delta(t-t'), 
\label{eqn:ortho_wt}
\\
\int_\realone\dd{t}e^{\pm i(\omega-\omega')t} &= 2\pi\delta(\omega-\omega'),
\label{eqn:ortho_tw}
\\
\int_{\realone^3}\dd[3]{k}e^{\pm i\bk\cdot(\br-\br')} &= \left(2\pi\right)^3\delta^{(3)}(\br-\br'), 
\label{eqn:ortho_kr}
\\
\int_{\realone^3}\dd[3]{r}e^{\pm i(\bk-\bk')\cdot\br} &= \left(2\pi\right)^3\delta^{(3)}(\bk-\bk').
\label{eqn:ortho_rk}
\end{align}

\subsubsection{Orthogonality relations for lattice waves}
I will make frequent use of the following orthogonality
relations.
\begin{align}
\frac{1}{N_c}\sum_\bR e^{i(\bk-\bk')\cdot\bR} &= \delta_{\bk\bk'}
\label{eqn:ortho1}
\\
\frac{1}{N_c}\sum_\bk e^{i\bk\cdot(\bR-\bR')}&=2\delta_{\bR\bR'}
\label{eqn:ortho2}
\\
\frac{1}{N_c}\sum_{\{\bk,\bbk\}} e^{i\bk\cdot(\bR-\bR')}&=\delta_{\bR\bR'}.
\label{eqn:ortho3}
\end{align}

\section{Real and complex vector spaces} 
\label{section:vectors}
The decision to work with complex vectors and exponentials, instead
of real vectors and real sinusoids, simplifies calculations
and
makes it possible to work with enough generality to describe, simultaneously, many
possible sets of boundary conditions. 
However, it slightly complicates any operation 
that involves the use of a metric tensor, such as projecting one vector onto another 
or calculating a vector's norm. 

\subsection{Vector notation, inner products, metrics, and dual vectors}
\label{section:vectornotation}
In a Euclidean vector space of arbitrary dimension, with basis vectors ${\be_\alpha}$, the inner product
of two vectors ${\bv=\sum_{\alpha}v^\alpha\be_\alpha}$ and ${\bw=\sum_{\alpha}w^\alpha\be_\alpha}$ is 
\begin{align*}
\bv\cdot\bw &\equiv g(\bv,\bw) = g(\bw,\bv)
=\sum_{\alpha\beta} v^\alpha g(\be_\alpha,\be_\beta) w^\beta 
\end{align*}
where 
$g$ is the Euclidean metric, which is symmetric and bilinear and often denoted by a dot, i.e., 
${g(\be_\alpha,\be_\beta)=\be_\alpha\cdot\be_\beta}$.
However, when we double the dimension of a Euclidean space by extending each of its
dimensions into the complex plane, we must use a different metric, $\eta$, for the inner product.

For example, the complex plane $\complex$ can be viewed as a two dimensional vector space
over the set of real numbers, $\realone$. However, this space is only isometrically isomorphic 
to the Euclidean space, ${\realone^2\equiv\realone\times\realone}$, if the 
symmetric bilinear
metric, ${\eta(a,b)\equiv \frac{1}{2}\left(a^* b + a b^*\right)= \eta^*(a,b)=\eta(b,a)}$, 
is used in $\complex$. 
It is only with this metric that 
the expression for ${a}$ can be expressed in a new basis, ${\{u,v\}}$,
as ${a = \eta(u,a)u+\eta(v,a)v}$ and the norm of $a$ can be expressed as
${\abs{a} =\sqrt{\eta(a,a)}}$.

Similarly, the inner product of two vectors ${\bv, \bw\in \complex^n}$
is the $n$-dimensional version of this inner product ${\eta(\bv,\bw)=\eta^*(\bv,\bw)=\eta(\bw,\bv)}$, which is related
to the Euclidean dot product as ${\eta(\bv,\bw)= \frac{1}{2}\left[g(\bv^*,\bw)+g(\bw^*,\bv)\right]}$.

I will use the symbols $\eta$ and $g$ to denote
the complex symmetric metric and the Euclidean metric, respectively, in spaces of all dimensions.
I do this with the understanding that they always denote the metrics of the space 
that their vector arguments belong to.

I will continue to use
boldface type (e.g., $\bu$) to denote vectors in $\realone^3$ or $\complex^3$, but
I will use
$\ket{u}$ to denote a vector in ${\realone^{3N}}$ or ${\complex^{3N}}$
and ${\kket{u}}$ to denote a vector in ${\realone^{3NN_c}}$ or ${\complex^{3NN_c}}$.

I will denote the duals of the vectors ${\bu}$, ${\ket{u}}$, 
and ${\kket{u}}$ by ${\bu^\dagger}$, ${\bra{u}}$, and ${\bbra{u}}$,
respectively, where the {\em dual}, or {\em metric dual}, of a vector is the unique linear map
from the vector space to its field of scalars ($\realone$ or $\complex$)
provided by the inner product. For example, in ${\complex^3}$ we have
${\bu^\dagger:\complex^3\to\complex; \bv\mapsto \bu^\dagger\bv\equiv g(\bu^*,\bv)}$, 
where ${\bu^*}$ is ${\bu}$ after each taking the complex conjugate of each of its
components and the basis vectors in which it is expressed. The difference
between ${\bu^*}$ and ${\bu^\dagger}$ is that ${\bu^*}$ is a vector and
${\bu^\dagger}$ is the operator ${\bu^*\mathbf{\cdot}\;\equiv g(\bu^*,\;)}$, i.e., ${\bu^\dagger\bv=\bu^*\cdot\bv}$.

In ${\complex^{3N}}$ the dual of ${\ket{u}}$ is 
${\bra{u}:\complex^{3N}\to\complex; \ket{v}\mapsto \mel{u}{}{v}\equiv \braket{u}{v}}$;
and in ${\complex^{3NN_c}}$ we have
${\bbra{u}:\complex^{3NN_c}\to\complex; \kket{v}\mapsto \bbra{u}\kket{v}\equiv \bbraket{u}{v}}$.
I will sometimes use $\dagger$ to denote the dualizing operation in higher dimensions;
for example ${\ket{u}^\dagger\equiv \bra{u}}$. 

I will denote the inner product of
two vectors in ${\realone^3}$ by 
${\bu\cdot\bv\equiv g(\bu,\bv)}$ 
and in ${\complex^3}$ I will denote the inner product by
${\eta(\bu,\bv)= \frac{1}{2}\left[g(\bu^*,\bv)+g(\bv^*,\bu)\right]}$.
In spaces of dimensions ${3N}$ and ${3NN_c}$ I
denote the inner products by ${\eta(\ket{u},\ket{v})}$ and ${\eta(\kket{u},\kket{v})}$, respectively, 
and I will also use a hybrid of an inner 
product and a {\em Clifford product}~\cite{hestenes,doran}, 
denoted by ${\braket{u}{v}\equiv g(\ket{u}^*,\ket{v})}$
and ${\bbraket{u}{v}\equiv g(\kket{u}^*,\kket{v})}$, respectively. 
The real parts of ${\braket{u}{v}}$ and ${\bbraket{u}{v}}$ are inner products
of vectors in the real vector spaces ${\realone^{6N}}$ and ${\realone^{6NN_c}}$, 
respectively, but the meanings of their imaginary parts are less easy to interpret.
I briefly discuss this further in App.~\ref{section:clifford_product}.

\subsection{Clifford product}
\label{section:clifford_product}
A Clifford product of two vectors is the sum of their inner and outer products. The inner product
is a scalar and the outer product is a bivector. For example, when working in $\realone^2$ with 
orthonormal basis ${\{\be_1,\be_2\}}$, the Clifford product of ${\bu\equiv u^1\be_1 + u^2\be_2}$
and ${\bv\equiv v^1\be_1+v^2\be_2}$ is
\begin{align*}
\bu\bv & = \bu\cdot\bv + \bu\wedge\bv 
\\
&= \left(u^1v^1 +u^2v^2\right) + \be_1\be_2\left(u^1v^2-u^2 v^1\right),
\end{align*}
where the Clifford product of bivector ${\be_1\be_2}$ with itself is ${-1}$.
Knowledge of Clifford algebra is not necessary to understand the present work, but \onlinecite{hestenes,doran}
are two of many good starting points for readers interested in learning about it. 

For the purposes of
this work, it suffices to know that, when complex numbers are used, and when
I perform multiplications like
\begin{align*}
u^* v&\equiv (u^1-iu^2)(v^1+iv^2)
\\
&=(u^1v^1+u^2v^2)+i(u^1v^2-u^2 v^1), 
\end{align*}
I am implicitly taking a Clifford product. Therefore ${i(u^1v^2-u^2 v^2)}$ can be interpreted, geometrically, as a bivector
whose magnitude is the area of the parallelogram with edges ${\bu}$ and ${\bv}$ and
which is parallel to a plane containing both $\bu$ and $\bv$.
That plane has two sides, one of which is parallel to ${\bu\wedge\bv}$ and antiparallel to ${\bv\wedge\bu}$, 
and the other of which is parallel to ${\bv\wedge\bu}$ and antiparallel to ${\bu\wedge\bv}$. Which is which is a
matter of convention and will not concern us.

\subsection{Dual vectors in spaces of dimension $3$, ${3N}$, and ${3NN_c}$}
\label{section:dualvectors}
In ${\complex^3}$ an arbitrary vector $\bv_1$ and its metric dual are denoted as 
\begin{align*}
\bv_1\equiv\sum_\alpha v_1^\alpha\,\ba_\alpha \longleftrightarrow 
\bv_1^\dagger\equiv \sum_{\alpha\beta}v_1^{\alpha *}g_{\alpha\beta} \bb^\beta.
\end{align*}
In ${\complex^{3N}}$ an arbitrary vector ${\ket{v_2}}$ and its metric dual are denoted
as
\begin{align*}
\ket{v_2}\equiv\sum_{j\alpha} v_2^{j\alpha}\,\ket{j\alpha} \longleftrightarrow
\bra{v_2}\equiv \sum_{j \alpha\beta}v_2^{j \alpha *}g_{\alpha\beta} \bra{j\beta}.
\end{align*}
In ${\complex^{3 N N_c}}$ an arbitrary vector ${\kket{v_3}}$ and its metric dual are denoted
as
\begin{align*}
\kket{v_3}\equiv\sum_{\bR j\alpha} v_3^{\bR j\alpha}\,\kket{\bR j\alpha}
\longleftrightarrow \bbra{v_3}\equiv \sum_{\bR j \alpha\beta}v_3^{\bR j \alpha *}g_{\alpha\beta} \bbra{\bR j\beta}.
\end{align*}
It follows that the square moduli of these vectors are 
\begin{align*}
\bv_1^\dagger\bv_1 & = g(\bv_1^*,\bv_1) = \sum_{\alpha\beta} v_1^{\alpha *} g_{\alpha\beta} v_1^\beta
\\
\braket{v_2} & = g(\ket{v_2}^*,\ket{v_2})= \sum_{j\alpha\beta}v_2^{j\alpha *}g_{\alpha\beta}v_2^{j\beta}
\\
\bbraket{v_3}{v_3} & = g(\kket{v_3}^*,\kket{v_3})=\sum_{\bR j\alpha\beta}v_3^{\bR j\alpha *}g_{\alpha\beta}v_3^{\bR j\beta} 
\end{align*}
The Clifford product of a vector with itself equals its inner product with itself, because the outer
product of a vector with itself vanishes.

\subsection{The imaginary part of an inner product in complex vector spaces}
\label{section:complex_inner_product}
At the end of App.~\ref{section:vectornotation} I referred to ${\braket{u}{v}}$ and ${\bbraket{u}{v}}$ 
as hybrids of inner and Clifford
products. By this I mean, for example, that if the orthnormal basis ${\{\ket{e_{j \alpha}}\}}$ spans ${\realone^{3N}}$, 
and therefore also spans ${\complex^{3N}}$, then any vector ${\ket{u}}$ can be
expressed as ${\ket{u}=\sum_\alpha u^{j \alpha}\ket{e_{j \alpha}}}$, for some set of coordinates 
${u^{ j \alpha}\in\complex}$.
Then,
\begin{align}
\braket{u}{v}
&=\left(\sum_{ j \alpha} u^{ j \alpha }\ket{e_{j \alpha}}\right)^\dagger
\left(\sum_{ i \beta} v^{ i \beta}\ket{e_{ i \beta}}\right)
\label{eqn:clifford1}
\\
	&=\sum_{j} \sum_{\alpha\beta} u^{j \alpha *} \delta_{\alpha\beta}v^{j \beta} 
	 =\sum_{j\alpha}  u^{j *}_\alpha v^{j \alpha},
\nonumber
\end{align}
where ${u^j_\alpha=u^{j\alpha}}$ because the basis is orthonormal.
If ${\braket{u}{v}}$ was a true Clifford product, it would have a bivector
part consisting of a sum of terms proportional to the outer products ${\ket{e_{j\alpha}}\wedge\ket{e_{i\beta}}}$
of different basis vectors. However, we take the inner product 
${\braket{e_{j\alpha}}{e_{i\beta}}=\delta_{ij}\delta_{\alpha\beta}}$ of the basis vectors and, implicitly, 
by using complex numbers, we find the sum ${u^{j\alpha *}v^{j\alpha}}$ of the inner and outer products of the 
{\em components} along these basis vectors. Then we add all these sums of inner and outer products together.
Therefore the real part of the result is a sum of real inner products, which is an inner product of vectors
in a vector space of dimension ${6N}$. 
The imaginary part is difficult to interpret because the imaginary unit ${i=\sqrt{-1}}$
plays the role of a bivector whenever ${\complex}$ is used as a substitute for ${\realone^2}$, 
but when ${j\alpha\neq i\beta}$, the products ${u^{j \alpha *}v^{j \alpha}}$ 
and ${u^{i\beta *}v^{i \beta}}$ are products of pairs of
vectors that belong to {\em different} two dimensional vector spaces, 
identified by ${j\alpha}$ and ${i\beta}$, respectively.
Therefore, the imaginary part of ${\braket{u}{v}}$ appears to be nonsensical.

However, let us momentarily abandon our notational conventions for vectors
in spaces of dimension ${3N}$ and $3$, and let us 
express ${\ket{e_{j\alpha}}}$ as the tensor product
${\ket{e_{j\alpha}}=\ket{j}\otimes\ket{\alpha}}$, 
where ${\ket{j}\in\{0,1\}^N}$, ${\ket{\alpha}\in\realone^3}$, 
and the orthogonality relations ${\braket{i}{j}=\delta_{ij}}$
and ${\braket{\alpha}{\beta}=\delta_{\alpha\beta}}$ hold, and imply the
relation,
\begin{align*}
\braket{e_{j\alpha}}{e_{i\beta}}=\left(\bra{\alpha}\otimes\bra{j}\right)\left(\ket{i}\otimes\ket{\beta}\right)
=\braket{i}{j}\braket{\alpha}{\beta}=\delta_{ij}\delta_{\alpha\beta}.
\end{align*}
Then, for each term proportional to ${\braket{e_{j\alpha}}{e_{i\beta}}=\delta_{ij}\braket{\alpha}{\beta}}$ 
in Eq.~\ref{eqn:clifford1}, 
there is a term proportional to ${\braket{e_{i\beta}}{e_{j\alpha}}=\delta_{ij}\braket{\beta}{\alpha}}$.
Putting it another way, for every imaginary term that represents a bivector parallel to ${\ket{\alpha}\wedge\ket{\beta}}$, 
there is an imaginary term of equal magnitude and opposite sign that represents 
a bivector parallel to ${\ket{\beta}\wedge\ket{\alpha}=-\ket{\alpha}\wedge\ket{\beta}}$.
These terms cancel one another in the nonsensical sum.

\vspace{1cm}
\onecolumngrid
\bibliography{phononpaper}
\end{document}